\documentclass[a4paper,fleqn,usenatbib]{mnras}

\usepackage{newtxtext,newtxmath}

\usepackage[T1]{fontenc}
\usepackage{ae,aecompl}

\usepackage[utf8]{inputenc}  
\usepackage{graphicx}	

\title[Clear signs of accretion in the Sextans dSph]{Appearances can be deceiving: clear signs of accretion in the seemingly ordinary Sextans dSph}

\author[L. Cicuéndez \& G. Battaglia]{
	L. Cicuéndez,$^{1,2}$\thanks{E-mail: lcicuend@iac.es}
	and G. Battaglia,$^{1,2}$
	\\
	$^{1}$Instituto de Astrofísica de Canarias (IAC), C/Vía Láctea, s/n, 38205, San Cristóbal de la Laguna, Tenerife, Spain\\
	$^{2}$Departamento de Astrofí­sica, Universidad de La Laguna, 38206, San Cristóbal de la Laguna, Tenerife, Spain
}

\date{Accepted 2018 June 28. Received 2018 June 6; in original form 2018 April 6}

\pubyear{2018}

\begin{document}
	\label{firstpage}
	\pagerange{\pageref{firstpage}--\pageref{lastpage}}
	\maketitle
	
	\begin{abstract}
        We report the discovery of clear observational signs of past accretion/merger events in one of the Milky Way satellite galaxies, the Sextans dwarf spheroidal (dSph). These were uncovered in the spatial distribution, internal kinematics and metallicity properties of Sextans stars using literature CTIO/DECam photometric and Magellan/MMFS spectroscopic catalogues. We find the spatial distribution of stars to vary as a function of the colour/metallicity, being rather regular and round for the blue  (metal-poor) red giant branch and main-sequence turn-off stars but much more elliptical and irregularly shaped for the red (metal-rich) ones, with a distinct ``shell-like'' overdensity in the northeast side. We also detect kinematic anomalies, in the form of a ``ring-like'' feature with a considerably larger systemic line-of-sight velocity and lower metallicity than the rest of stars; even the photometrically selected component with a regular looking spatial distribution displays complex kinematics. With a stellar mass of just $\sim5\times10^{5} M_{\odot}$, Sextans becomes the smallest galaxy presenting clear observational signs of accretion to date.
	\end{abstract}
	
	\begin{keywords}
		galaxies: individual: Sextans dSph -- galaxies: dwarf -- galaxies: structure -- galaxies: evolution -- Local Group -- dark matter		
	\end{keywords}
	
	
	\section{Introduction}
	
	According to the $\Lambda$-Cold Dark Matter ($\Lambda$CDM) cosmological framework of structure formation, Milky Way (MW)-like and massive galaxies are formed in a hierarchical process through mergers and accretion of smaller systems \citep{white78}. Nevertheless, mergers are also expected between low mass haloes \citep[e.g.][]{fakhouri10}. \cite{deason14} showed, using dissipationless cosmological zoom-in simulations, that very few isolated dwarf galaxies in the Local Group (LG) could have escaped from a major merger in their formation history. More recently, also \cite{benitez-llambay16}, using cosmological zoom-in simulations of the LG formation, argued that mergers played a significant role in the formation of at least some isolated dwarfs. Regarding satellite dwarf galaxies, these are half as likely to have experienced a major merger as compared to isolated dwarfs of similar mass \citep{deason14}.
	
	From an observational perspective, different substructures have been discovered in the kinematics and density distribution of the stellar component of several LG dwarf galaxies, probably resulting from the disruption of smaller accreted dwarfs. Among the galaxies with a clear observational evidence of external accretion, the smallest one up to date is And~II ($\sim10^{7}$ solar masses in stars), in which \cite{amorisco14} detected the presence of a stellar stream. Subsequently, several studies modeled the galaxy as the result of a merger between two dwarfs \citep[see e.g.][]{ebrova15,fouquet17}, reproducing also And~II observed prolate rotation \citep{ho12}. Recently, prolate rotation was also detected in the Phoenix transition-type dwarf; this kinematic feature, combined with the peculiar spatial distribution of the young stars, was interpreted as a sign of an accretion or merger event \citep{kacharov17}. Another complex system is the Fornax dSph: \cite{battaglia06} found a double peaked line-of-sight (LOS) velocity distribution for the metal-poor stars of this galaxy, which, together with the asymmetric and more centrally concentrated distribution of its young stars, could have resulted from the recent accretion of external material after the merger with another dwarf, consistent with a previous evidence of a shell-like feature \citep{coleman04} and the scenario later proposed by \cite{amorisco12} where Fornax formed from a late merger of a bound pair.
	
	Other examples of substructures not necessarily formed by accretion of smaller galactic systems are also present among the dwarf galaxies satellites of the Milky Way. For example, Ursa Minor shows a secondary peak in its stellar density \citep{bellazzini02}, where \cite{kleyna03} then detected a cold kinematic signature also confirmed by \cite{pace14}. In both these latter works it was argued that this cold kinematic sub-structure would probably correspond to a stellar object that survived to multiple orbits within the gravitational potential of the dwarf galaxy thanks to, either having its own dark matter halo protecting it, or due to Ursa Minor dark matter halo being cored instead of cuspy. The Sculptor dwarf spheroidal galaxy also presents cold kinematic substructures \citep[see Chapter 4 in][]{battaglia07}. As for the Sextans dSph, \cite{kleyna04}, \cite{walker06} and \cite{battaglia11} (hereafter K04, W06 and B11 respectively) identified localized cold kinematic substructures in the inner regions of this system; thanks to the availability of metallicity estimates for the individual stars, B11 traced the signature back as belonging to Sextans' metal-poor component.

    In this work we report the detection of clear observational signs of accretion in the Sextans dSph, which impact on both the spatial distribution and internal kinematics of the dwarf galaxy. This paper is structured as follows: in Sect.~\ref{sec:photometry} we use the \cite{yo18} (hereafter C18) CTIO/DECam photometric catalogue and show that the spatial distribution of blue RGB (bRGB) and main-sequence turn-off (MSTO) stars is remarkably different from the one of red RGB (rRGB) and MSTO stars; in Sect.~\ref{sec:kinematics} we re-analyze the MMFS spectroscopic sample from \cite{walker09} (hereafter W09) with the membership probabilities derived in C18 and detect the presence of a ``ring-like'' feature with a distinct velocity from the rest of Sextans's population and having a lower relative metallicity. Finally, we summarize the results and discuss the possible origins of the detected features in Sect.~\ref{sec:con}.  
	
	\section{Anomalous differences between the spatial distribution of blue and red RGB stars}
	\label{sec:photometry}
	
	The CTIO/DECam photometric dataset published by C18	consists of a catalogue of point-like sources with SDSS $g-$ and $r-$ magnitudes covering nearly 20 deg$^2$ around the Sextans dSph and reaching down to $\sim2$ mag below Sextans' oldest main-sequence turn-off. 
	
	Here we use this data-set to analyze the spatial distribution of Sextans stars as a function of their colour. The majority of the following analysis is based on sources brighter than $g=23$, because this guarantees that the photometric catalogue is unaffected by artificial overdensities appearing in the outer parts of the DECam pointings (see C18); at the distance of Sextans, this magnitude cut essentially corresponds to the base of the RGB. We will however also use stars in the range $23<g<24$, that is on the MSTO, down to the $S/N\sim10$ level of the shallowest pointing in the mosaic, to verify that the results hold also for the stars in this evolutionary phase.  
	
\begin{figure}
		\includegraphics[trim={2.75cm 0cm 3.75cm 1.5cm},clip,width=\hsize]{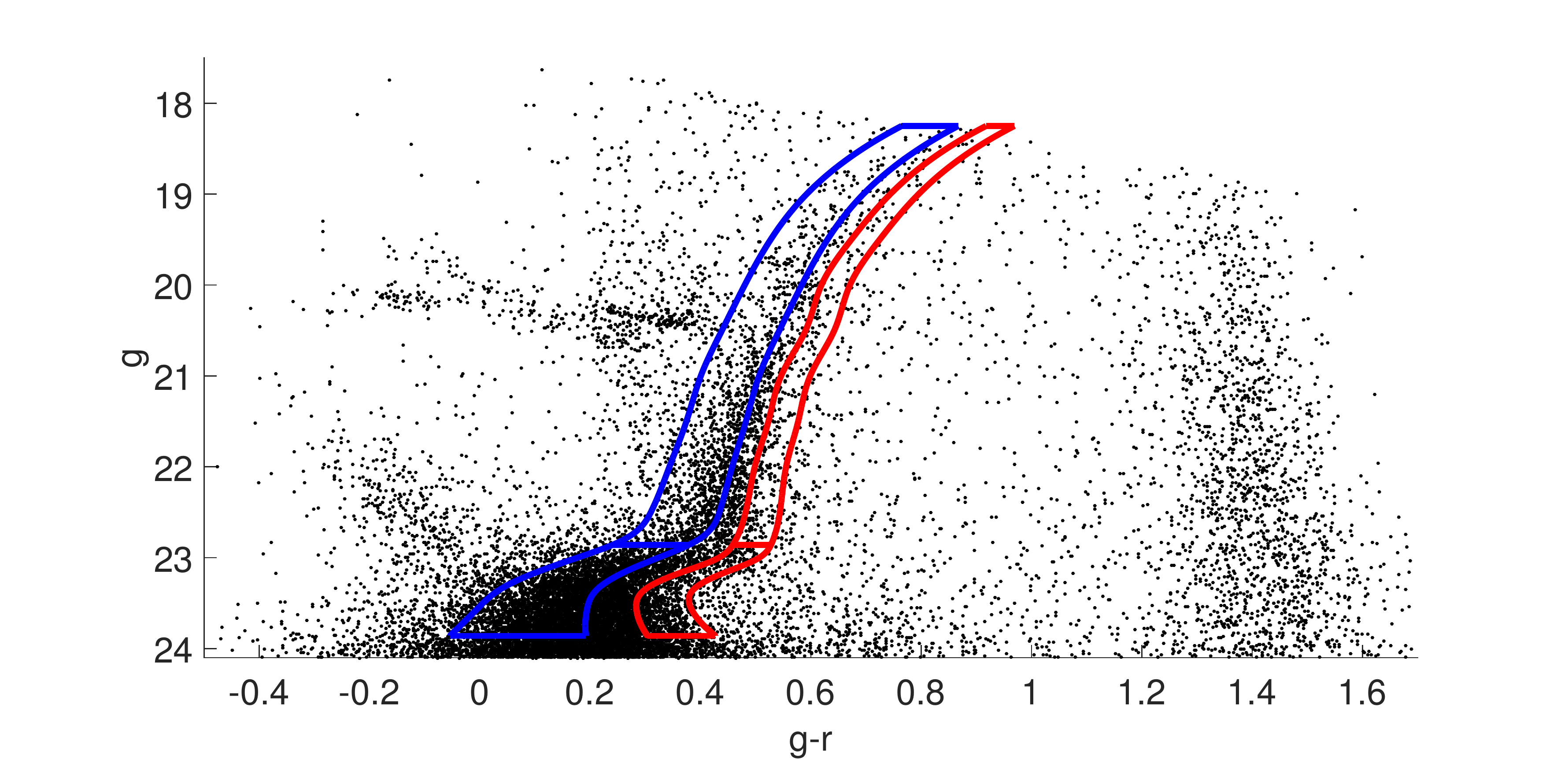}	
		\caption{CMD from C18 CTIO/DECam dataset for stars with major axis distance less than 0.5 deg from Sextans centre. Blue and red lines show the selection masks used to select blue and red RGB (and MSTO) stars, respectively. The horizontal line at $g=23$ separates the RGB from the MSTO selection (see main text).}
		\label{fig:cmd}
	\end{figure}

	The colour-magnitude diagram (CMD) selection regions adopted for the analysis are displayed in Fig.~\ref{fig:cmd}. To define the CMD masks we first selected stars within a major axis distance of half a degree, in order to avoid most of the contaminants, and then retained only point-like sources compatible with being MSTO, sub-giants or RGB stars. The resulting CMD was then binned along the magnitude axis ($g$) and the 85th and 50th percentiles of the colour distribution in a given magnitude bin were evaluated; finally, these selection limits were cubic spline interpolated at each data point of the whole dataset, so that each object could be then easily assigned either to the blue or to the red selection. We used the 85th percentile of the colour distribution to define the ``blue'' (inner) limit of the red selection box, and the 50th percentile for the ``red'' limit of the blue selection, while their outer limits are defined by these ones plus one and minus two times the median colour uncertainties at each magnitude bin, respectively. Note that each selection box likely contains a fraction of stars that originally belong to the other one due to the colour uncertainties, which are typically larger than the gap between both selection boxes. 
	
\begin{figure*}
		\includegraphics[trim={8.75cm 0cm 9cm 1.25cm},clip,width=\columnwidth]{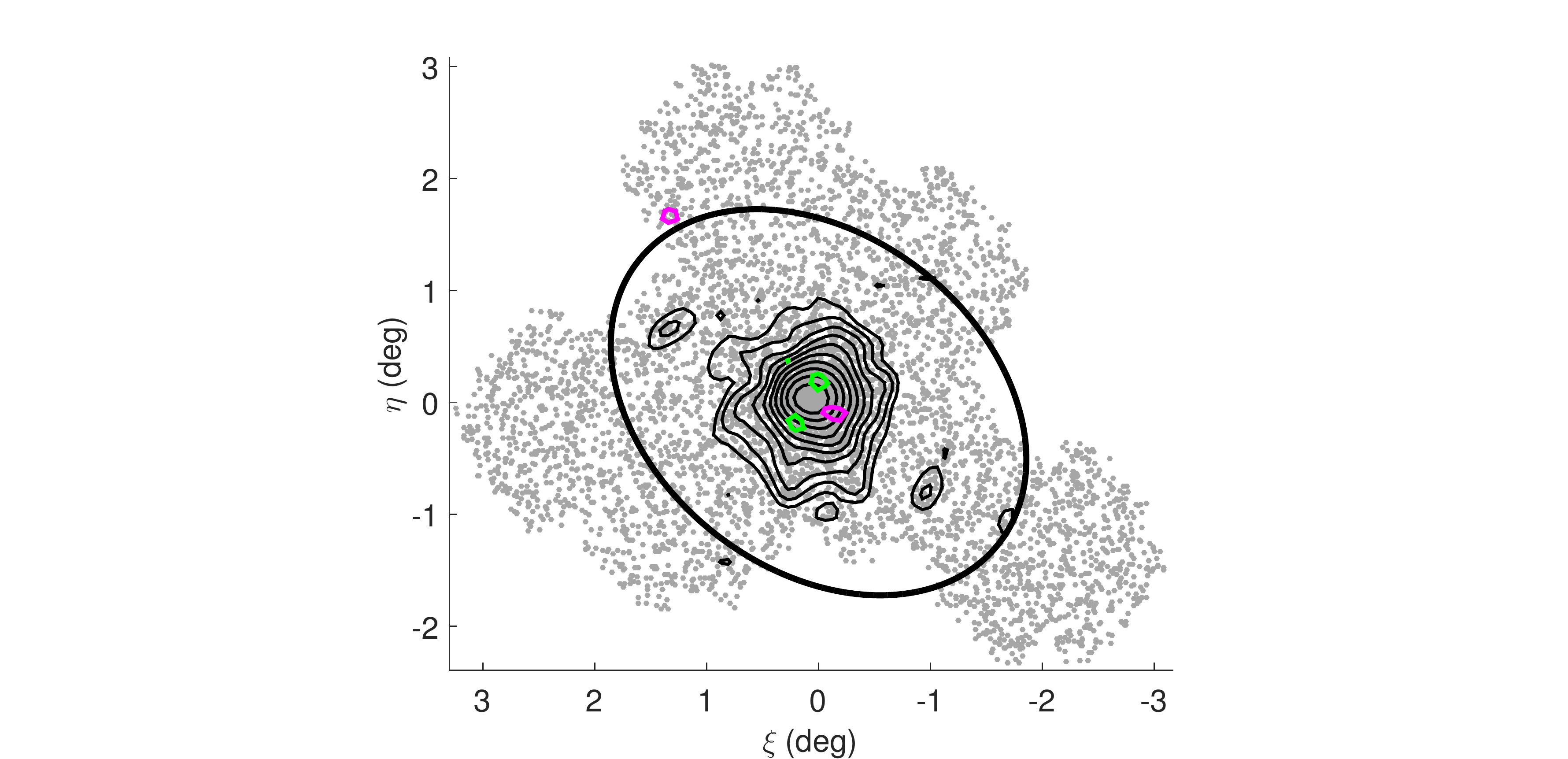}		
		\includegraphics[trim={8.75cm 0cm 9cm 1.25cm},clip,width=\columnwidth]{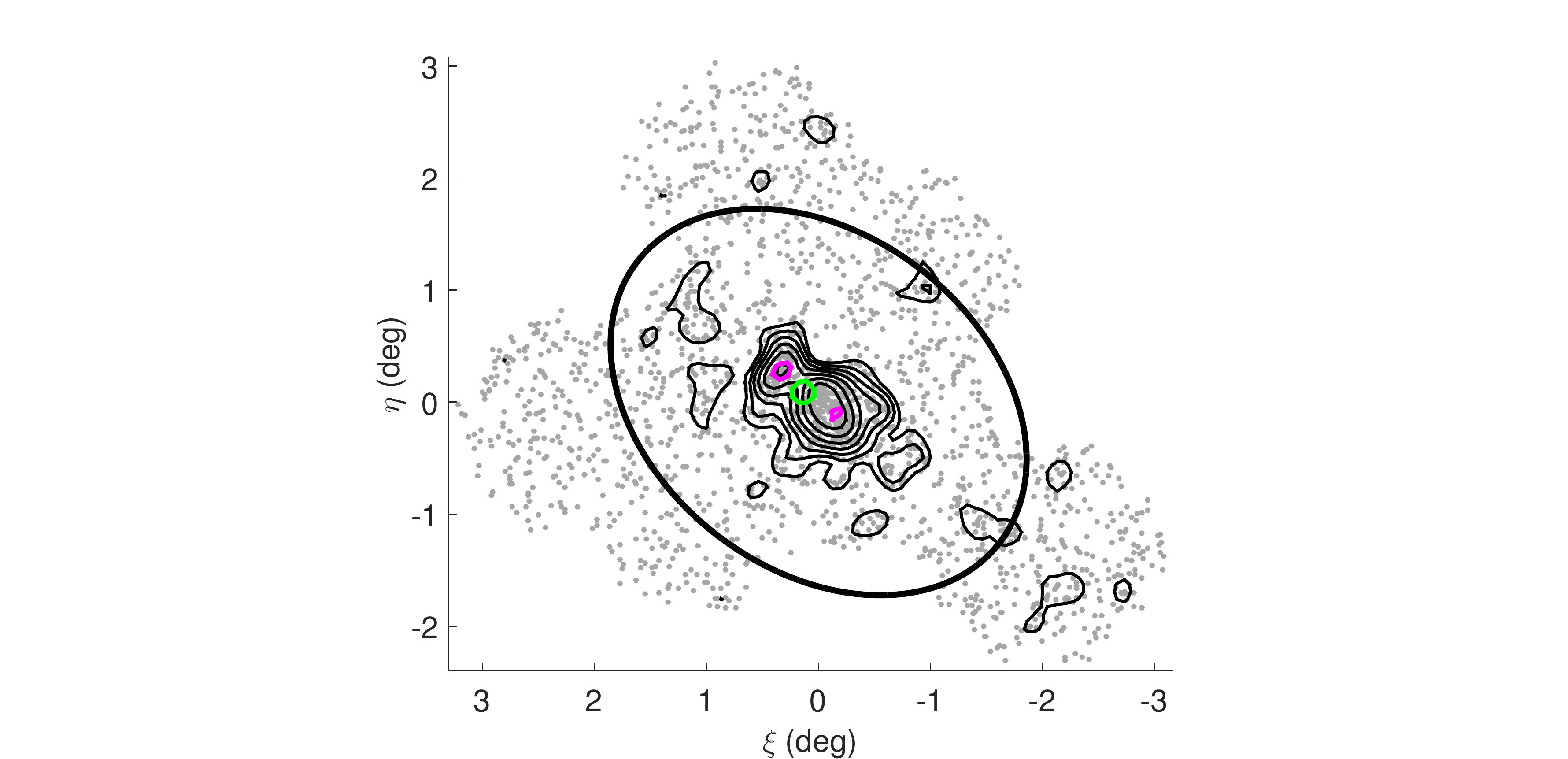}
		\includegraphics[trim={8.75cm 0cm 9cm 1.25cm},clip,width=\columnwidth]{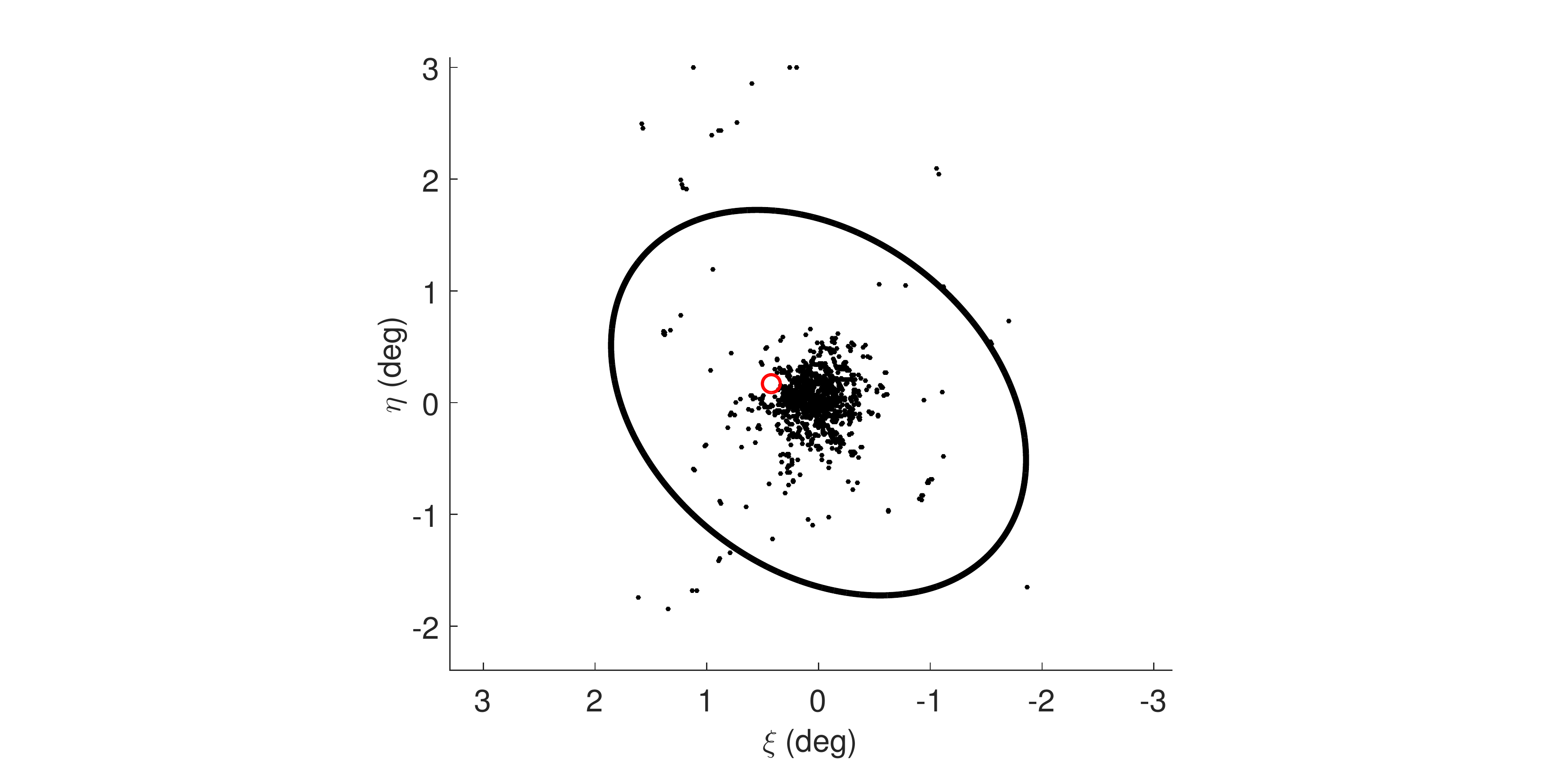}
		\includegraphics[trim={8.75cm 0cm 9cm 1.25cm},clip,width=\columnwidth]{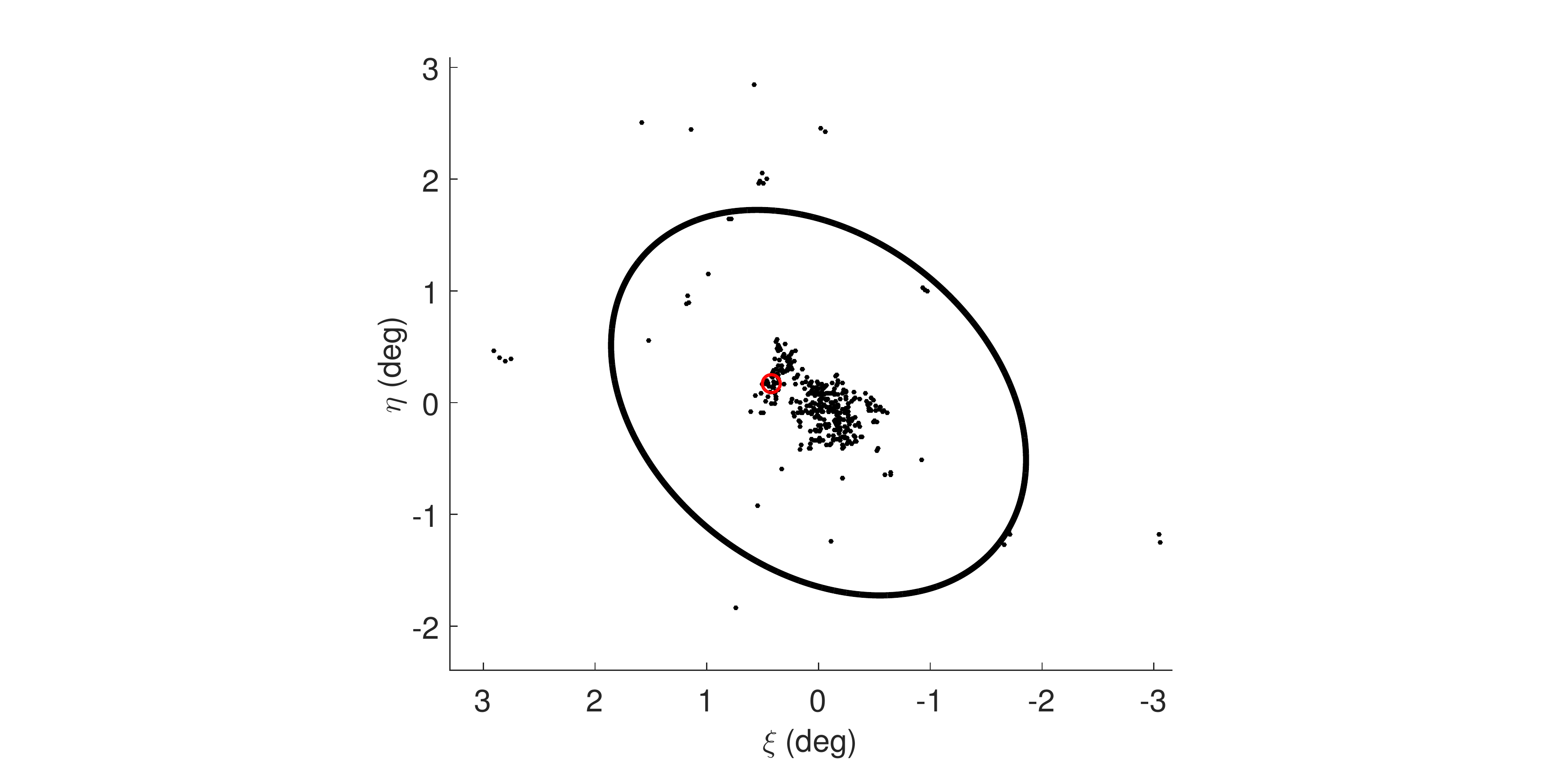}
			
		\caption{Top: Spatial distribution of blue (left) and red (right) RGB stars (see CMD masks in Fig.~\ref{fig:cmd}), with overlaid iso-density contours from the matched-filtering method denoting exponentially increasing values of stellar number density, with the outermost contours tracing the 3$\sigma$ detection. Green and magenta contours respectively trace the 3$\sigma$ detections above and below the mean of the residuals between the decontaminated matched-filtered maps of the corresponding deep CMD selection boxes (MSTO) and the mean of their best-fitting axisymmetric models (King, Sersic, Plummer and Exponential profiles). Black ellipses show the nominal King tidal radius derived in C18. Bottom:  Spatial distribution of blue (left) and red (right) RGB stars after decontaminating the stellar locations with the method explained in the Appendix. Red circles delimit the region centered on (0.42,0.17) deg where a considerably higher concentration of redder stars was detected (see also Fig.~\ref{fig:superclump}).}
		\label{fig:maps}
	\end{figure*}

	Figure~\ref{fig:maps} shows the spatial distribution of bRGB and rRGB point-like sources. These have been decontaminated following two different methods: a matched-filter analysis as in C18 (see iso-density contours in the top-panels) and a decontamination method acting on the individual point-sources without spatial binning (bottom panels), described in the Appendix. One can appreciate a rather regular and almost round spatial distribution of the bRGB stars, while the spatial distribution of rRGB stars is clearly more elliptical and with an irregular shape, with the presence of an overdensity of stars in the northeast side, reminiscent of the shells observed in much larger galaxies, and the centre of the bulk of the population slightly displaced southwest with respect to the one of the bRGB stars. We note that we have explored different color cuts and converged on the one that best highlights the differences in the spatial distribution of bRGB and rRGB stars, while not compromising number statistics.
	
	Being the spatial distribution of the bRGB fairly axisymmetric, we derived its structural parameters following the method by C18 based on \cite{richardson11}, which works by evaluating at the locations of the individual stars a likelihood expression which depends on the surface density profile to be fitted as well as on the ellipticity, position angle, etc. As in C18, in this expression we also included the surface density of contaminants by modelling it with a spatial bilinear distribution (i.e. a plane) to account for the expected spatial gradient in the density of MW contaminants. The most probable value of each structural parameter is then evaluated by sampling this likelihood expression through a Markov chain Monte Carlo (MCMC) sampler, where in our case we used ``The MCMC Hammer'' \citep{foreman-mackey13}\footnote{As in C18, the code we used (\url{https://github.com/grinsted/gwmcmc}) was developed at Centre for Ice and Climate (Niels Bohr Institute), for which we defined a total of 80 walkers, each of them doing approximately $10^4$ steps.}.
	
	The visual impression is confirmed by this quantitative analysis (see Tab.~\ref{tab:structural_parameters}): the ellipticity of the bRGB is 0.05, to be compared to the value of 0.27 of Sextans overall stellar population down to the same magnitude limit; the position angle of this blue RGB population is ill-determined, due the round shape; there is also a hint of a smaller half-light radius of the bRGB stars with respect to the overall Sextans stellar population, with the former having $\sim 19$ arcmin and the latter $\sim 22$ arcmin (see Table 3 of C18). We checked that the half-light radius of the bRGB population becomes even	smaller if adopting a bluer selection, reaching down to values as low as $\sim 17$ arcmin; this can probably be understood by the bRGB box containing stars originally belonging to the rRGB populations, scattered in the blue box by photometric errors in the colour, if the rRGB population has a larger half-light radius than the bRGB one (see below).	
	
	\begin{table}
		\caption{Structural parameters (median values of the marginalized posterior distributions) for Sextans stars falling in three of the selection boxes plotted in Fig.~\ref{fig:cmd}, derived with the MCMC Hammer when modelling the surface number density as an exponential profile. Regarding the 2D half-light radii $r_h$, they were derived from the exponential ones $r_e$ using $r_h=1.678\,r_e$ \citep{wolf10}. If we were to fit a Plummer profile, the only statistical difference would be in all the values of the 2D $r_{h}$ being $\sim 1-2$ arcmin larger.} 
		\centering                       					
		\begin{tabular}{c c c c}        
			\hline                
			\noalign{\smallskip}
			Parameter & Blue RGB & Blue MSTO & Red MSTO \\
			\noalign{\smallskip}
			\hline        
			\noalign{\smallskip}
			$\alpha_{2000}$ (º) & $153.282\substack{+0.008\\-0.008}$ & $153.285\substack{+0.004\\-0.004}$ & $153.25\substack{+0.02\\-0.02}$ \\     
			\noalign{\smallskip}
			$\delta_{2000}$ (º) & $-1.613\substack{+0.008\\-0.009}$ & $-1.619\substack{+0.004\\-0.004}$ & $-1.588\substack{+0.009\\-0.009}$ \\
			\noalign{\smallskip}
			Ellipticity & $0.05\substack{+0.05\\-0.04}$ & $0.05\substack{+0.03\\-0.03}$ & $0.44\substack{+0.04\\-0.04}$  \\
			\noalign{\smallskip}
			Pos. angle ($^\circ$) & $20\substack{+40\\-30}$ & $40\substack{+20\\-20}$ & $59\substack{+3\\-3}$ \\
			\noalign{\smallskip}
			$r_{e}$ ($'$) & $11.6\substack{+0.6\\-0.6}$ & $10.7\substack{+0.3\\-0.3}$ & $13.8\substack{+0.6\\-0.6}$ \\
			\noalign{\smallskip}
			2D $r_{h}$ ($'$) & $19\substack{+1\\-1}$ & $18.0\substack{+0.4\\-0.4}$ & $23\substack{+1\\-1}$ \\
			\noalign{\smallskip}
			\hline             
		\end{tabular}
		\label{tab:structural_parameters}      
	\end{table}
	
	\begin{figure}
		\centering
		\includegraphics[trim={3cm 0cm 9.5cm 1cm},clip,width=\hsize]{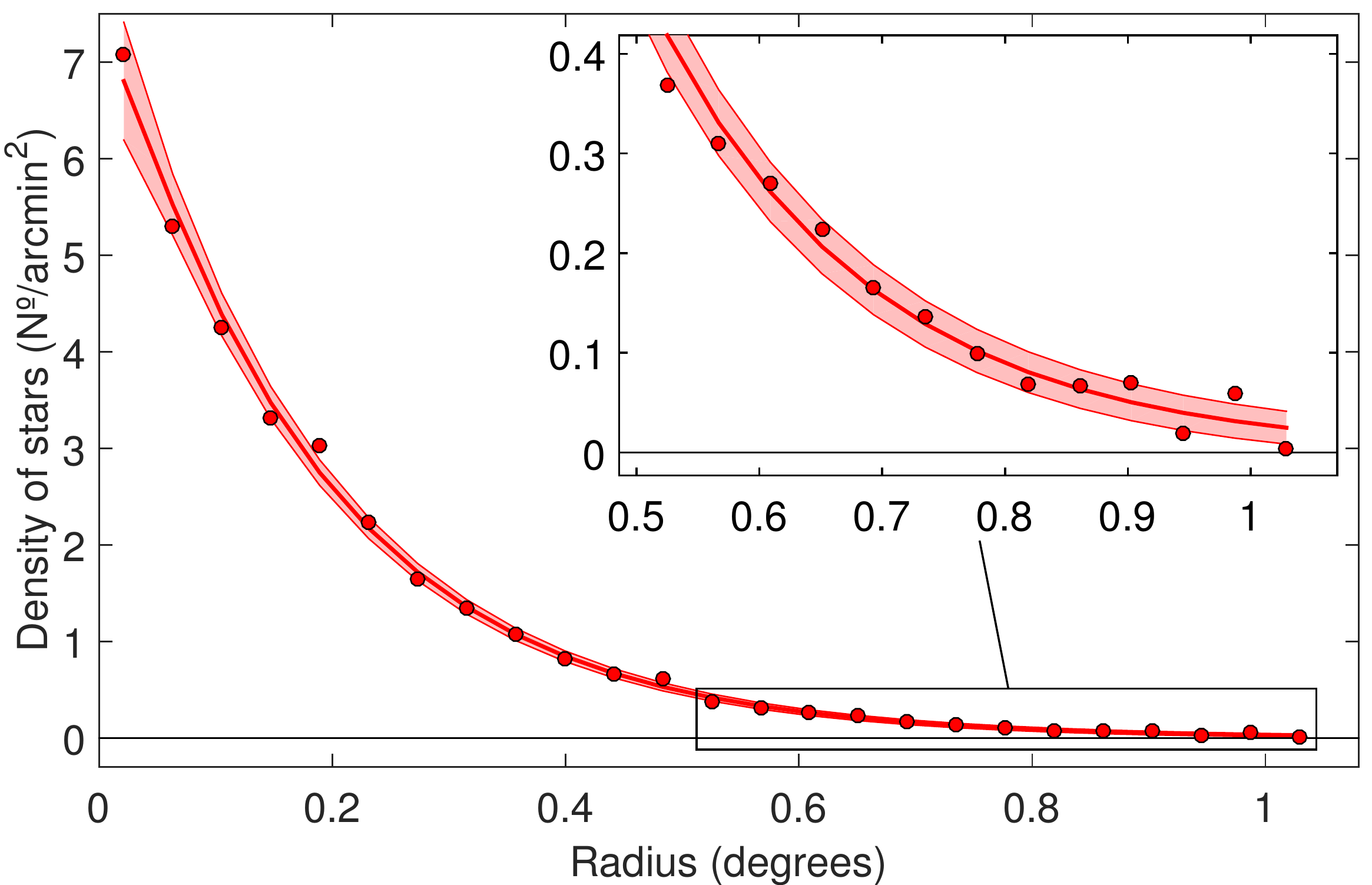}
		\caption{Contamination subtracted surface number density profile of the axisymmetric blue MSTO stars as a function of the major axis radius (with the external parts zoomed in), overlaid onto the $1\sigma$ confidence interval (red band) of the best-fitting exponential profile obtained with the MCMC Hammer. The $1\sigma$ confidence interval is computed from the best-fitting model assuming Poisson variances in each elliptical annulus.}
		\label{fig_fit_exp_blue_msto}
	\end{figure}
		
	Even though at magnitudes fainter than $g=23$ artificial artifacts starts appearing in the outer regions of the pointings, we repeated the analysis on the point-sources falling within the blue and red selection boxes in the magnitude range $23 < g < 24$ to test whether the features detected are preserved at fainter magnitudes, below the MSTO. In this case, we also derived the structural parameters of the red population,	which should be taken with a grain of salt due to the departures from axisymmetry in the spatial distribution of these stars. The results are summarized in Table~\ref{tab:structural_parameters}: the round shape of the blue population and much flatter distribution of the red one are confirmed, as it is the difference in the centre coordinates; furthermore, the red component is significantly less concentrated than the blue one, as reflected in the value of the  half-light radius. Since our aim here was not to determine which functional form best represents the surface number density of the stars in the different CMD selections, but to quantify relative differences in their spatial distribution, we report the results only for an exponential profile, in order to restrict the number of free parameters. Figure \ref{fig_fit_exp_blue_msto} shows that the best-fitting exponential model gives a satisfactory representation of the observed surface density profile of the axisymmetric blue MSTO, even despite the artificial artifacts in the outer regions of the pointings. We also tested whether the ``shell-like'' overdensity of rRGB stars in the northeast part of the field-of-view continues being present also in the fainter selection; to this end, we studied the distribution of residuals between the decontaminated matched-filter density map of the red stars from the deeper, MSTO selection and the best-fitting axisymmetric models: the 3$\sigma$ contours of over- and under-densities are plotted in the top panels of Fig.~\ref{fig:maps} and show that indeed the 3$\sigma$ over- and underdensities detected in the MSTO red selection coincide with the ``shell-like'' feature in the distribution of rRGB stars and the gap between such feature and the bulk of the rRGB population.	
	
	Using theoretical isochrones from the canonical models of the BASTI library\footnote{\url{http://albione.oa-teramo.inaf.it}} we constrained the mean metallicity of the blue and red selections to be within \mbox{$-3.6<$ [Fe/H] $<-2.6$} and \mbox{$-2.6<$ [Fe/H] $<-1.6$}, respectively. These two intervals are compatible with the two metallicity peaks found by B11 in the spectroscopic metallicity distribution function of the RGB stars, which the authors fitted with a sum of two Gaussians of means \mbox{[Fe/H] $=-2.7$} and \mbox{[Fe/H] $=-2.0$}. We also checked that these values are broadly in agreement with the spectroscopic metallicities measured by B11 for the stars falling in each of our RGB CMD selection boxes: the blue and red selections are indeed associated with B11 metal-poor and metal rich populations respectively and they are incompatible with being extracted from a common parent distribution in metallicity \mbox{(p-value $<$ 0.03} from a Kolmogorov-Smirnov test). 
	
	We conclude then that stars falling within the blue selection, which display a round and regular spatial distribution, encompass the metal-poor part of Sextans's stellar population, while those in the red selection, which have an asymmetric, clumpy and fairly	flat distribution, encompass the metal-rich population of Sextans. As we will see in Sect.~\ref{sec:kinematics}, anomalous properties can also be appreciated in the internal kinematics of Sextans's stellar component.  
	
	Here we notice that the presence of these stellar populations with distinctly different spatial distributions appears to be the culprit for the different structural parameters obtained by C18 as a function of magnitude when analyzing Sextans overall stellar population, where a deeper magnitude cut ($(g,r)=(24.9,24.9)$ Table D.1, C18) was yielding a rounder and more concentrated spatial distribution than when adopting a brighter cut ($(g,r)=(23.0,23.0)$, Table 3, C18). In that work the cause for this behavior could not be traced back to any obvious difference in the stellar population mix as a function of magnitude. Since at magnitudes fainter than $g=23$ the outer parts of the pointings of the C18 DECam dataset suffered from artificial overdensities caused by the morphological misclassification between extended and point-like objects due to out-of-focus regions, and fainter than $g\sim24$ small differences in depth between the pointings start appearing, C18 maintained a cautious approach and referred the reader to the structural parameters derived from the brighter portion of the photometric catalogue. Thanks to the analysis in this work, this behaviour can now be understood as due to a variable proportion of blue versus red stars as a function of magnitude, which we find to be higher at the deepest magnitudes. Therefore, in practice, the structural parameters derived for Sextans {\it are} dependent on the magnitude cut adopted, due to the varying proportion of blue/red stars having distinct spatial distributions. Strictly speaking, when regarding Sextans as a whole, the structural parameters derived from the deepest catalogue (Table~D.1, C18), which encompass more of Sextans's stars, should be more appropriate. 
		
	Finally, we note that the over- and under- densities detected in the middle panel of Fig.~7 in C18 in Sextans's overall stellar population can be confirmed as real features and understood as a consequence of the inadequacy of one single axisymmetric model in capturing the complexity of the underlying stellar population. 
	
\begin{figure}
		\includegraphics[trim={2.75cm 0cm 3.25cm 1.5cm},clip,width=\columnwidth]{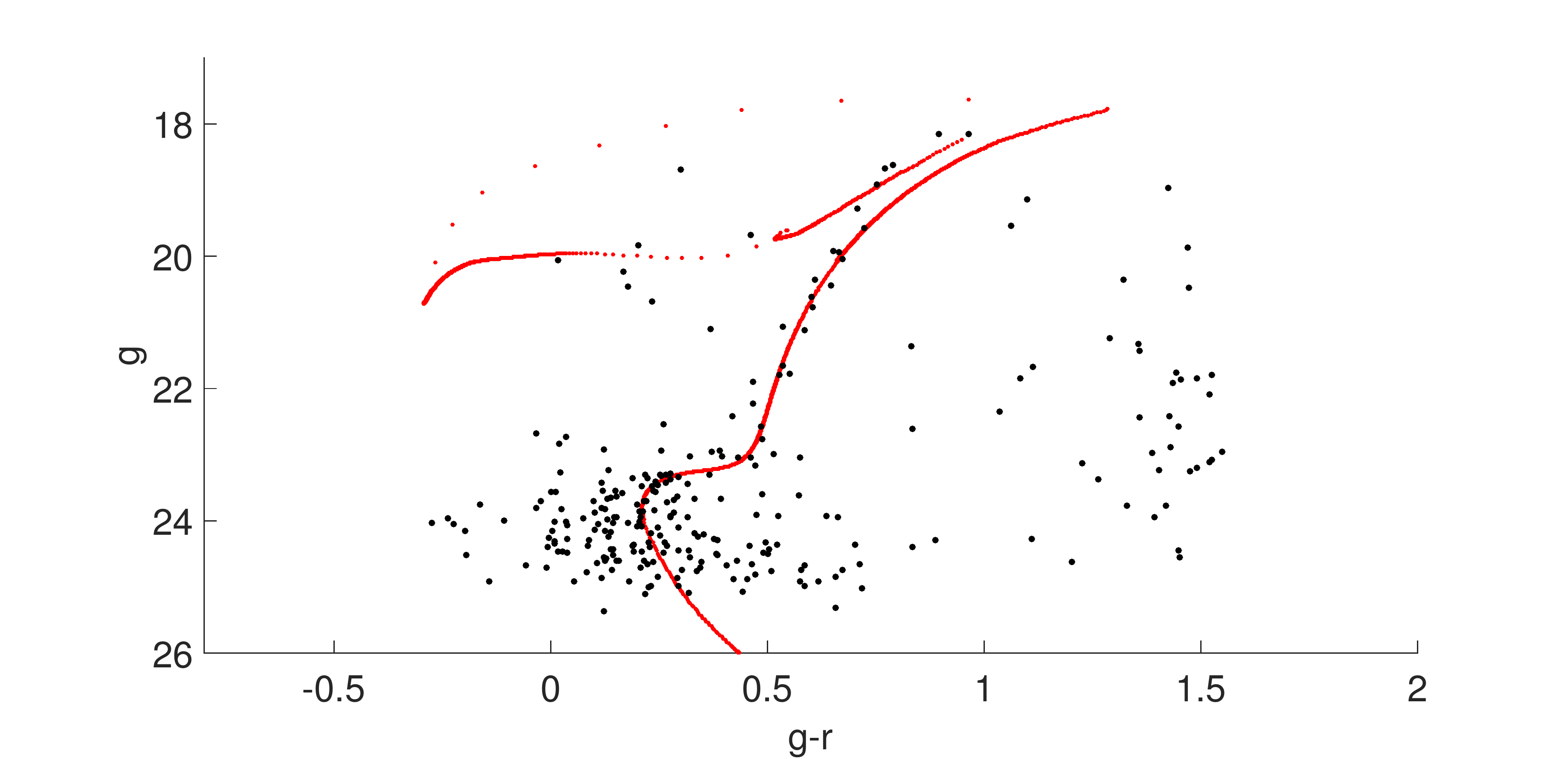}
		\includegraphics[trim={2.75cm 0cm 3.25cm 1.5cm},clip,width=\columnwidth]{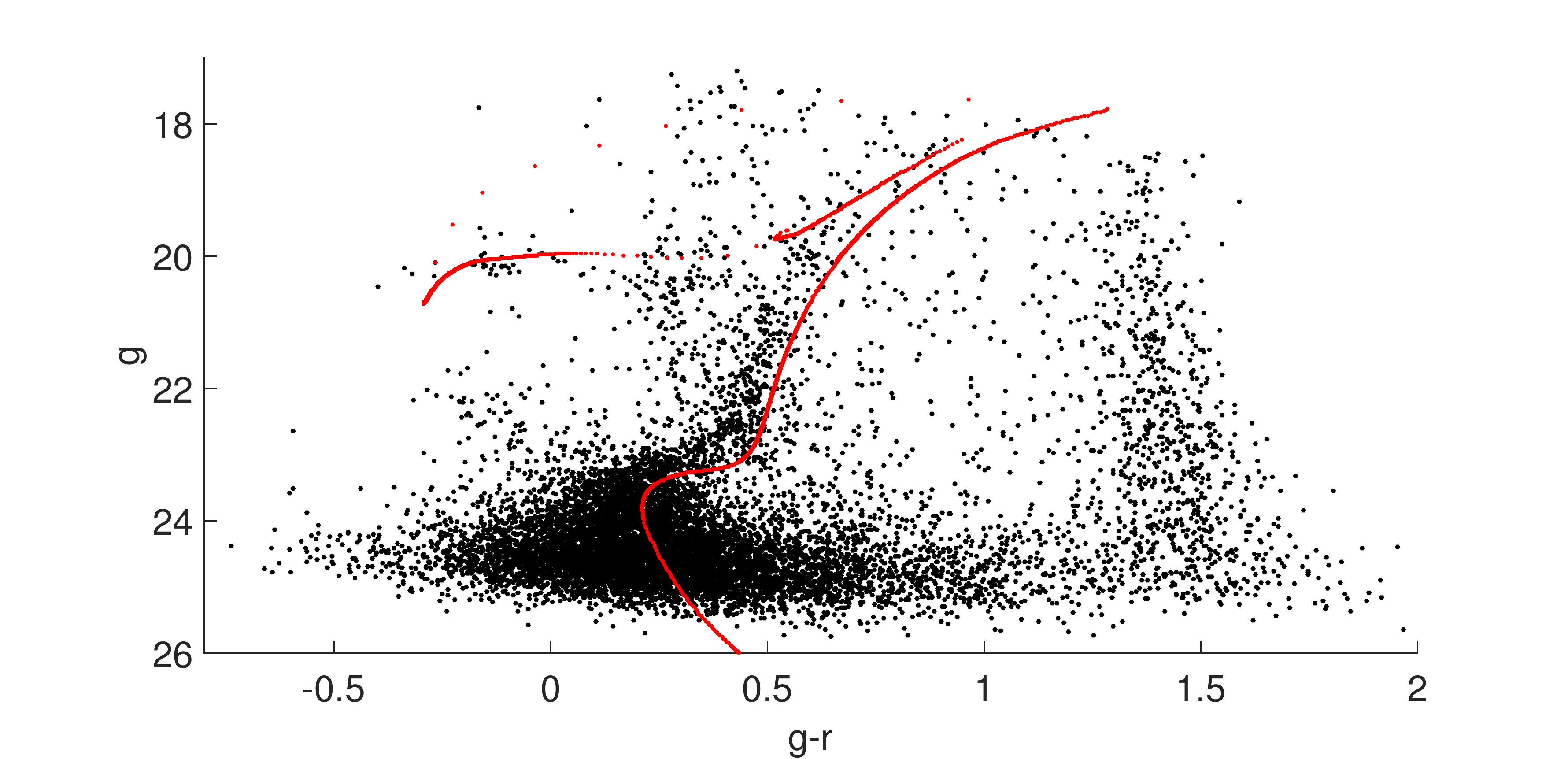}
		\caption{CMDs of the detected red blob at position (0.42,0.17) deg in the bottom panels of Fig.~\ref{fig:maps} (top) and of the rest of stars at the same major axis distance (bottom) for comparison; the isochrone (red line) has [Fe/H]$=-1.6$ and 13 Gyr in age and was produced by the canonical models of the BaSTI stellar library.}
		\label{fig:superclump}
	\end{figure}

	We also note that in the region marked with a red circle in the bottom panels of Fig.~\ref{fig:maps} the distribution of red stars is particularly clumpy, and coincides with an apparent decrease in the number of stars plotted in the decontaminated blue RGB map. No clear overdensity, such as a stellar cluster, is visible in the DECam images at that location. The CMD of this region is particularly red with respect to the	CMD of the remaining stars at the same elliptical radius (Fig.~\ref{fig:superclump}), and comparison with BaSTI isochrones suggests a metallicity [Fe/H]$=-1.6$, that is at the high metallicity end of what found for the red RGB and MSTO selection; this is perhaps signaling that star formation in that region occurred more recently than in the surroundings and could give hints as to the details of the process that caused the shell-like feature and asymmetric distribution of the red stars.
	
	\section{A ``ring-like'' kinematic substructure}
	\label{sec:kinematics}
	The W09 spectroscopic dataset consists of LOS heliocentric velocities and pseudo-equivalent widths of the Mg-triplet absorption feature ($\Sigma$Mg) obtained with the Michigan/MIKE Fiber System at the Magellan 6.5m Clay Telescope; here we include only those 440 RGB and horizontal branch (HB) stars with a probability higher than 95\% of being members to the Sextans dSph, as determined in C18.      
	
\begin{figure*}	
		\includegraphics[width=\columnwidth]{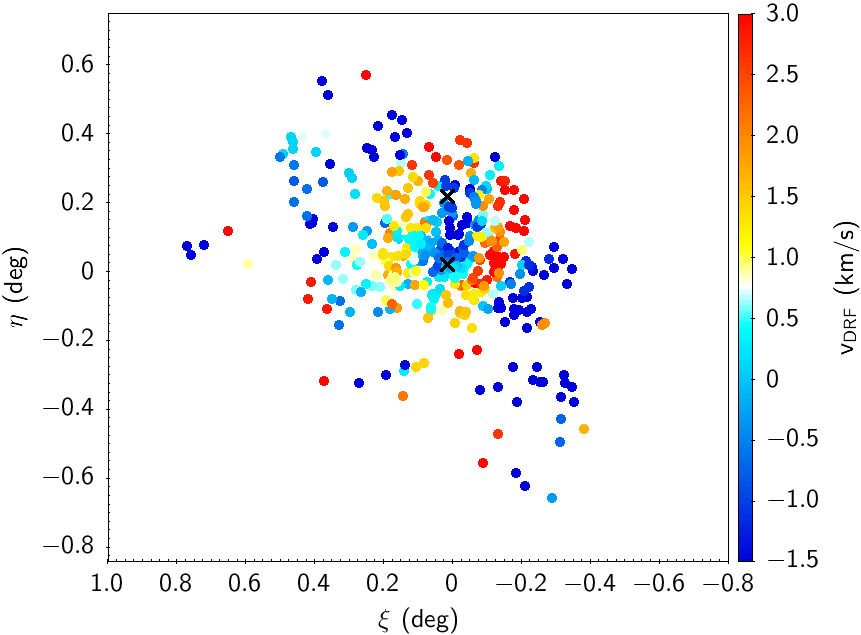}
		\includegraphics[width=\columnwidth]{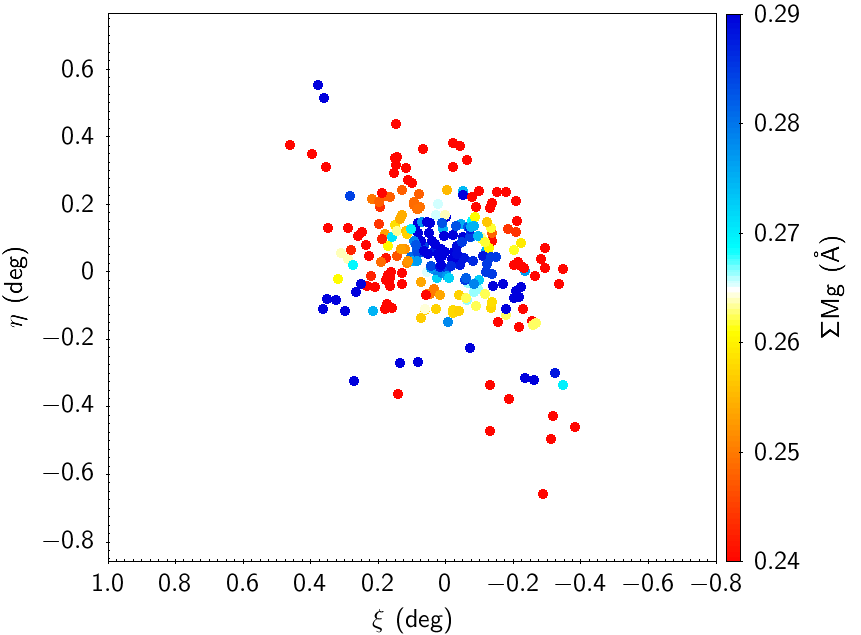}
		\includegraphics[width=\columnwidth]{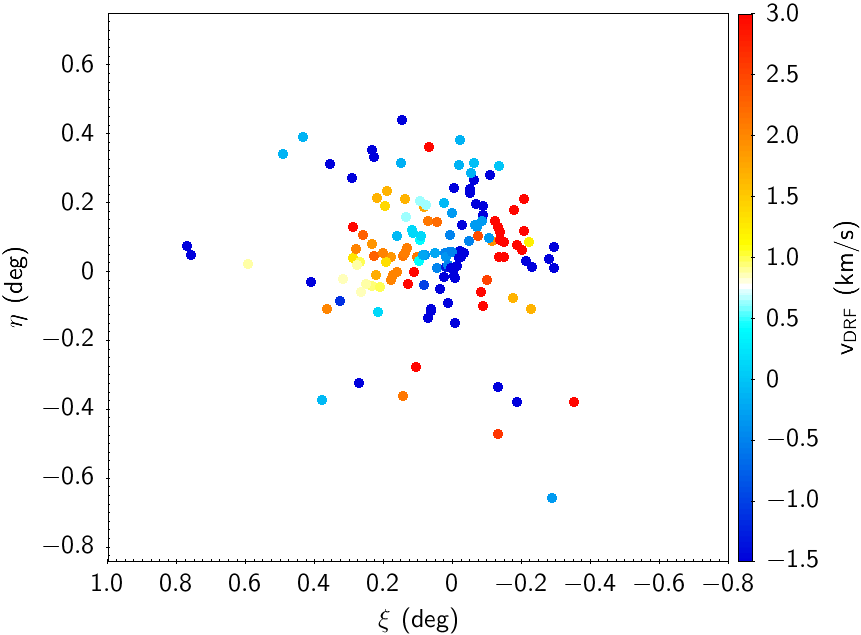}
		\includegraphics[width=\columnwidth]{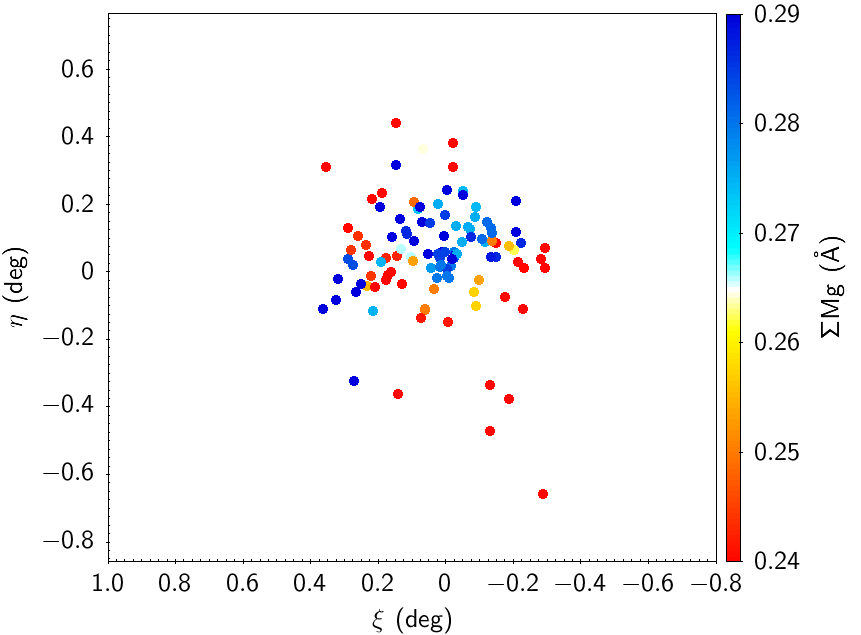}
		\includegraphics[width=\columnwidth]{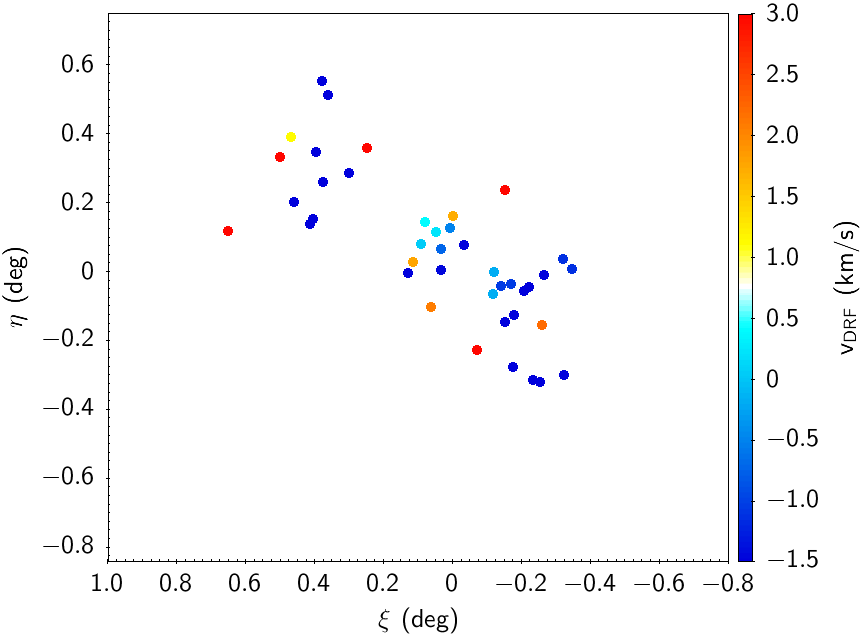}
		\includegraphics[width=\columnwidth]{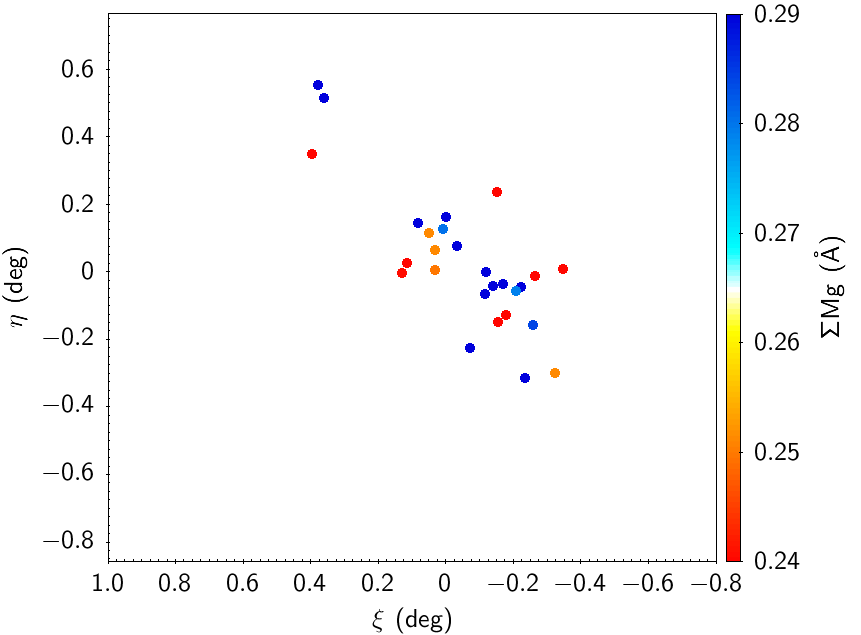}
		
		\caption{W09 spectroscopic sample of Sextans most probable members, colour-coded using median smoothed (with a kernel of $\sim 5$ arcmin) DRF LOS velocities (left) and $\Sigma$Mg values (right), with the latter excluding the HB stars. The middle and bottom panels correspond to the blue and red CMD selections plotted in Fig.~\ref{fig:cmd}, respectively. Those members whose values of DRF LOS velocities and $\Sigma$Mg are lower and higher than the range shown in the colourbar are plotted in blue and red. This was done to enhance the contrast and emphasize the ``ring'', given the fact that those stars with less close neighbors are not as smoothed as the central ones.
		Upper and lower black crosses in the top-left panel mark the centres of the cold kinematic substructures detected in W06, and in K04 and B11, respectively.}
		\label{fig:kine}	
	\end{figure*}

	The top-left panel of Fig.~\ref{fig:kine} shows the W09 LOS velocities with respect to the dwarf galaxy rest frame (DRF), after smoothing them with a kernel of $\sim 5$ arcmin. The original W09 LOS velocities in the heliocentric rest frame (HRF) were converted into the DRF ones using Eq.~4 of \cite{walker08}, a systemic velocity of 224.3 km/s (W09) and the transformations derived in their appendix, together with the HRF proper motion of Sextans dSph measured in \cite{casetti17}\footnote{After submission of the article to the journal, the Gaia collaboration released Gaia DR2 based systemic proper motions for the classical dwarf galaxies satellites of the Milky Way \citep{helmi18}. We have verified that using the Gaia collaboration systemic proper motion for Sextans to transform the velocities into the DRF system the results presented in this section are unaffected.}. In this plot we can distinguish a ``ring-like'' feature, which has a systemic LOS velocity larger than the rest of the stars; this feature is also visible when looking at the HRF velocities, before converting them into the DRF system. We note that we used only the W09 dataset because neither the VLT/FLAMES dataset by B11 nor the Keck/DEIMOS one by \cite{kirby10} have the appropriate spatial coverage to properly sample the detected feature; also, we decided not to combine datasets due to some systematics showing up in the comparison of velocities from overlapping Sextans members (see Figure A1 in B11), which could alter the signal. 
	
	Since the LOS velocities from the W09 Michigan/MIKE Fiber System dataset are the result of the weighted mean of measurements performed with the blue and red spectrograph channel, and sometimes of repeated measurements between overlapping pointings, we checked that the observed ``ring-like'' shape was not caused by any systematic difference between the velocities measured with one channel or the other, systematics in some fields or in the overlapping regions between them, or outliers in the individual measurements used to obtain the mean velocity of each star. We also analyzed the velocities of all the spectroscopic targets without considering their membership probabilities, searching for some foreground/background object in the LOS of Sextans that could be responsible of creating this ``ring-like'' feature on the kinematics, but none was found. We conclude then that the ``ring-like'' feature is real and belongs to the Sextans dSph.
	
	The ``ring-like'' feature is also clearly visible at the same location in a smoothed map of $\Sigma$Mg values\footnote{The $\Sigma$Mg values we use throughout this work were corrected for the dependence of opacity on the gravity and temperature of the star, which \cite{walker09c} showed correlates with V-V\textsubscript{HB} along the RGB. To this aim, we applied the correction in Eq.~3 of \cite{walker09c}, using the V magnitudes measured by W09 and B11. We excluded the stars compatible with being HB stars at the distance of	Sextans, because the correction was originally developed only for RGB stars.} (top-right panel of Fig.~\ref{fig:kine}), which also indicates that the stars that populate this feature have a lower metallicity relative to the rest.  
	
	In order to characterize this feature, we used the map of smoothed velocities as a reference to manually isolate the stars of the ``ring'' (152 sources) according to their spatial location and to define a control sample, that is the stars located in the inner and outer regions defined by the ``ring'' (179 sources). We note that to characterize these two samples we analyse the original measurements in velocity and $\Sigma$Mg, not the smoothed ones. To this end, before proceeding, we first performed a Kolmogorov-Smirnov (K-S) test on the measurements in DRF velocity and $\Sigma$Mg of the inner and outer regions of the control sample and verified that they are compatible with having been extracted from the same distributions in velocity and $\Sigma$Mg (\mbox{p-values $>$ 0.85} and 0.50, respectively); we can therefore treat them as a single control sample.	
		
\begin{figure*}
		\includegraphics[trim={1.9cm 0cm 2.65cm 1cm},clip,width=\columnwidth]{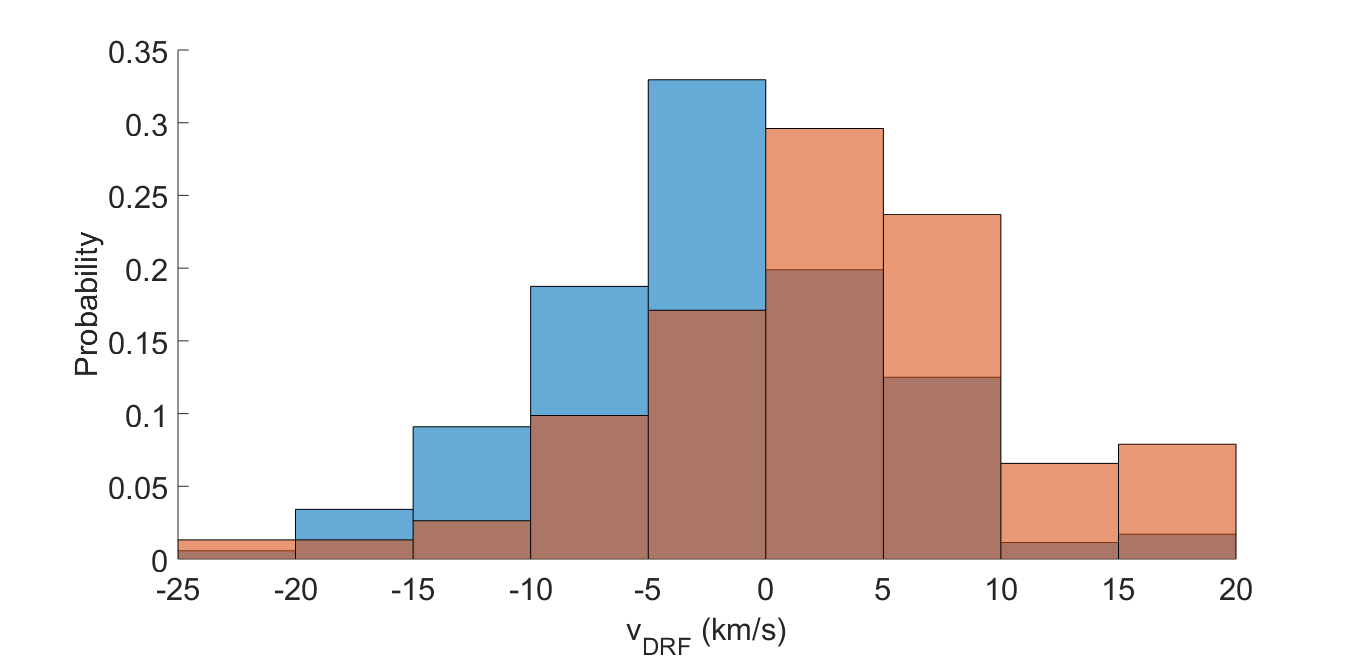}
		\includegraphics[trim={1.9cm 0cm 2.65cm 1cm},clip,width=\columnwidth]{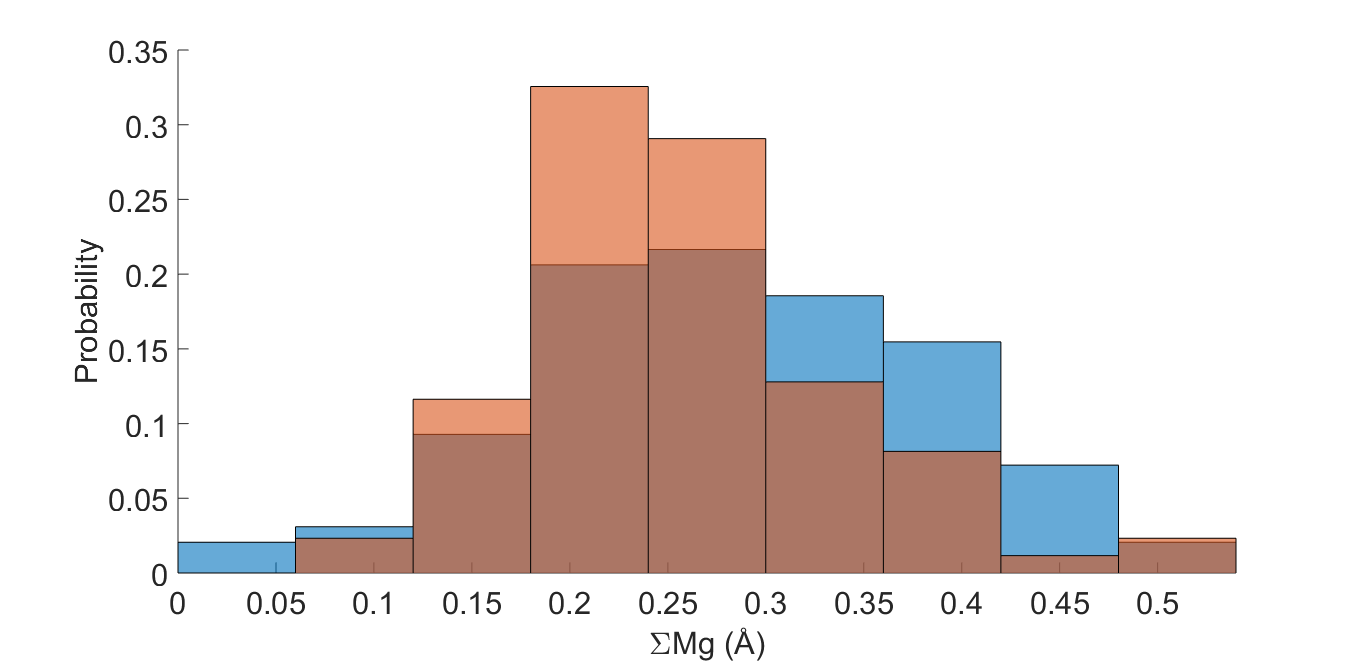}
		\includegraphics[trim={1.9cm 0cm 2.65cm 1cm},clip,width=\columnwidth]{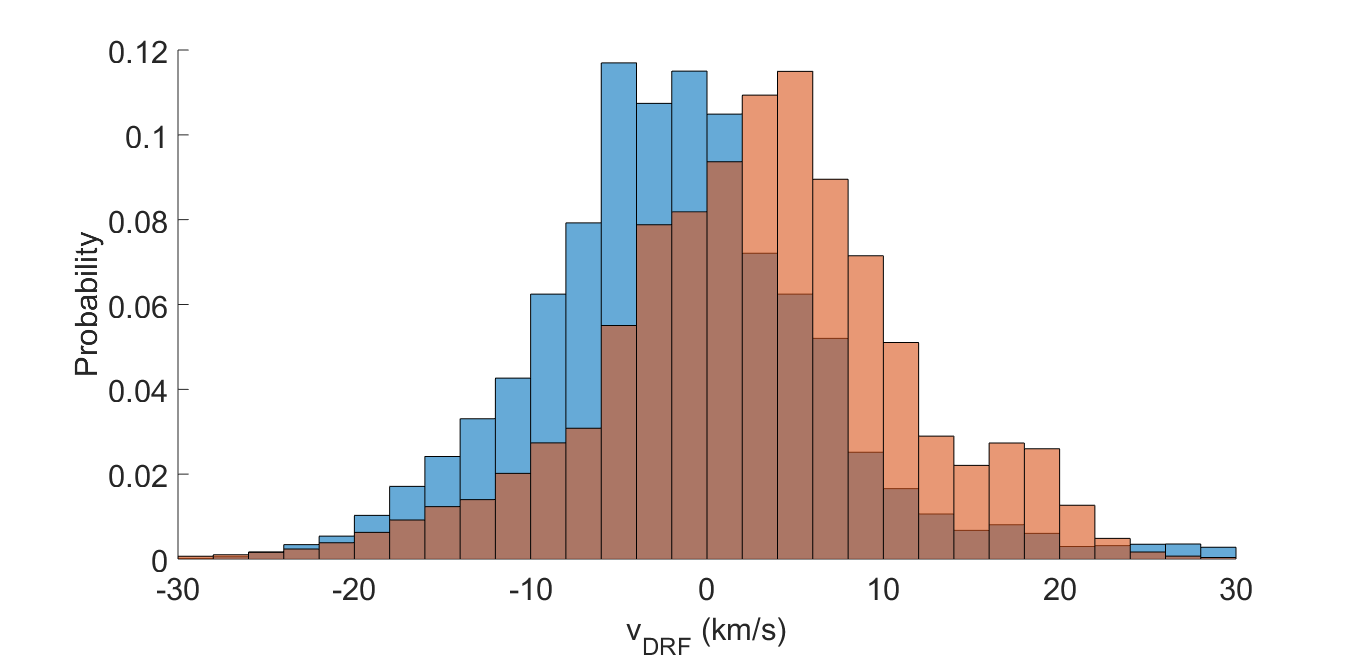}
		\includegraphics[trim={1.9cm 0cm 2.65cm 1cm},clip,width=\columnwidth]{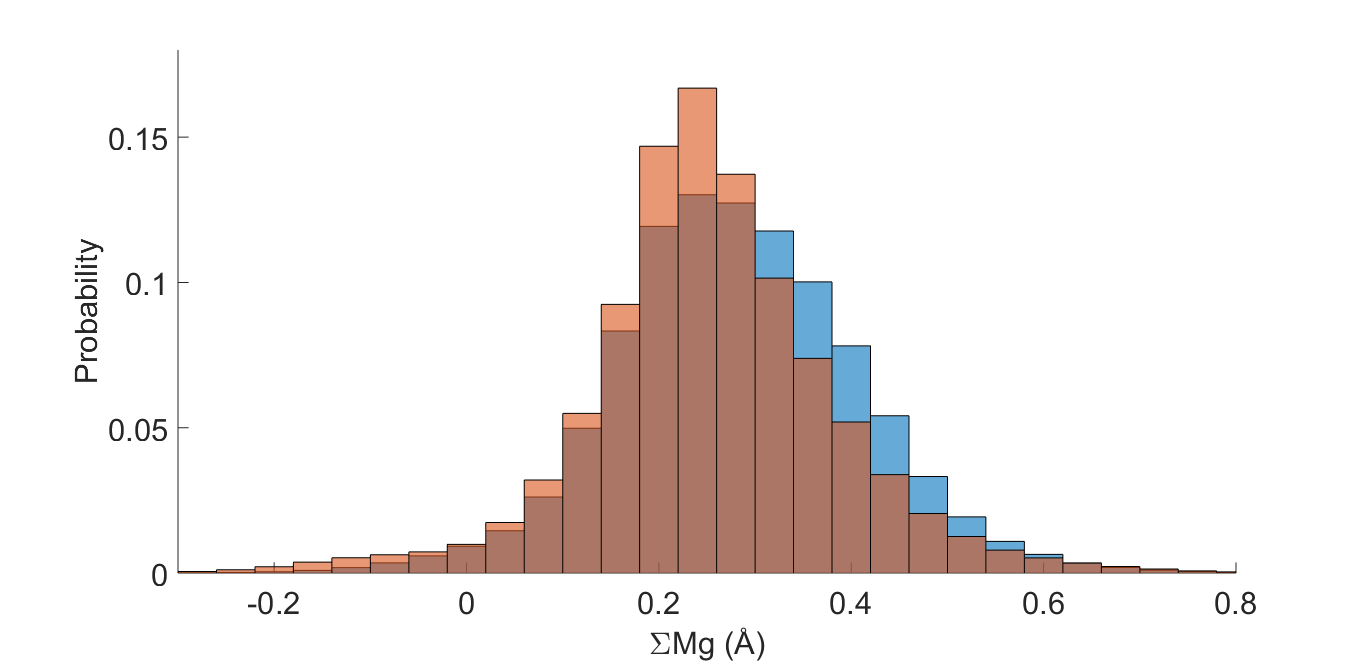}
		\includegraphics[trim={1.9cm 0cm 2.65cm 1cm},clip,width=\columnwidth]{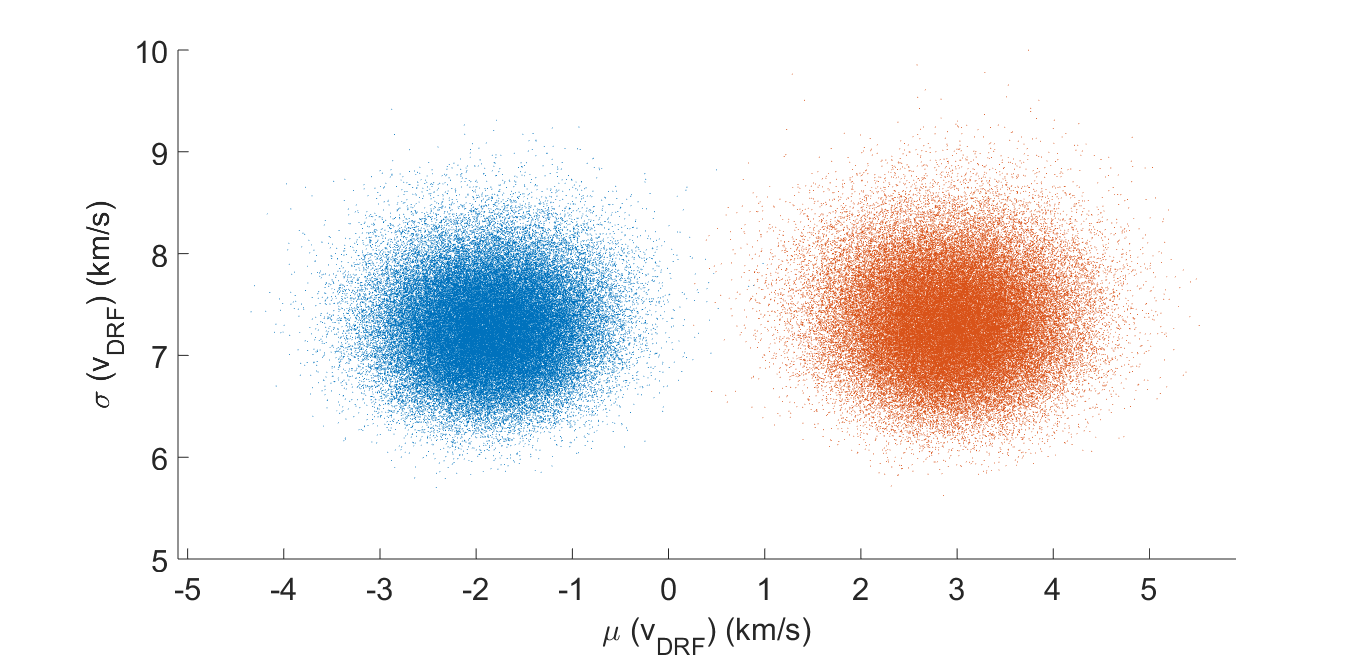}
		\includegraphics[trim={1.9cm 0cm 2.65cm 1cm},clip,width=\columnwidth]{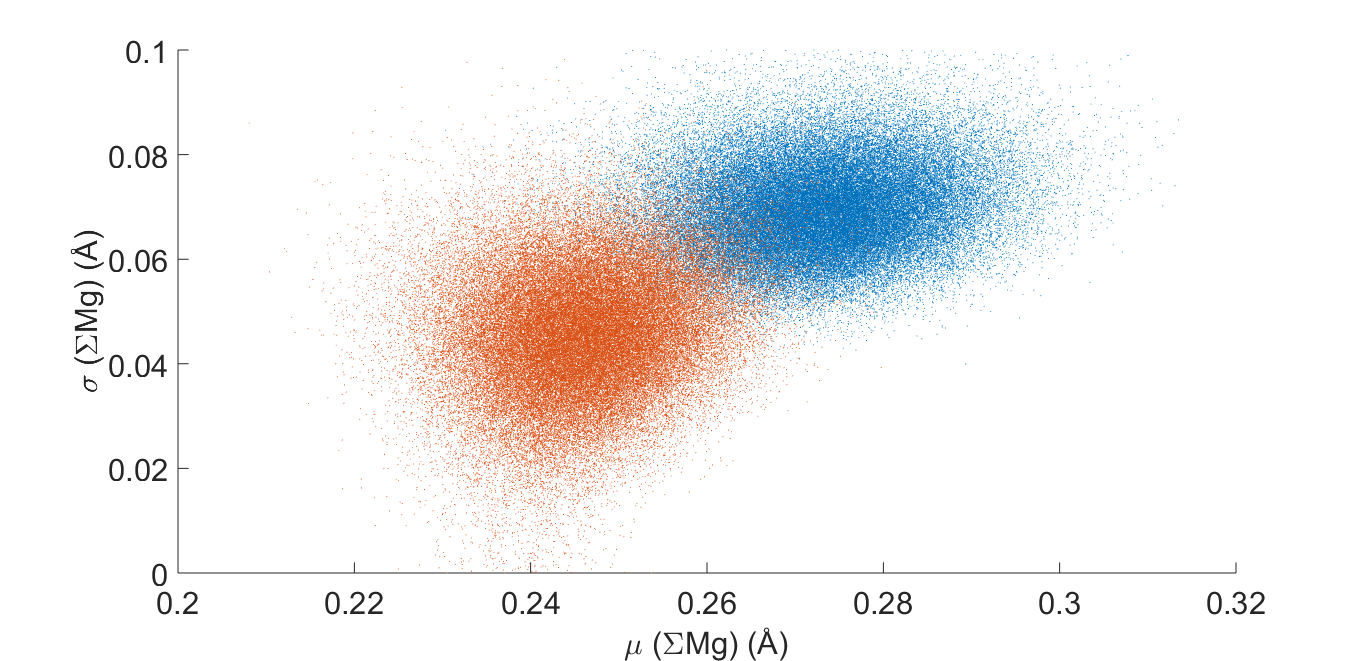}    
		\caption{Left: DRF LOS velocity analysis. Right: $\Sigma$Mg analysis. Top and middle panels: Distributions not convolved and convolved by the errors in the measurements, respectively. Bottom: Samplings from the MCMC Hammer of the mean and intrinsic standard deviation of the	distributions. The results corresponding to the ``ring'' and control samples are plotted in red and blue respectively.}
		\label{fig:distributions}
	\end{figure*}

	Already at a visual inspection, the distribution of the DRF velocities of stars in the region of the ``ring'' shows clear differences with respect to the control sample, and mild differences are detected also between the $\Sigma$Mg distributions (see Fig.~\ref{fig:distributions}), both when unconvolved (top) and convolved (middle) with the measurement errors;	in particular, when considering the unconvolved distributions, the mean velocities and $\Sigma$Mg confirm that the ``ring'' moves faster and is more metal-poor than the control sample, while no significant difference is measured in the spread (quantified by the standard deviation) for both observables (see Table~\ref{tab:stat_parameters}). If we approximate the distributions with Gaussians, as it is commonly done (an approximation valid to a first order only, as we will see below), and derive the mean and intrinsic dispersions via a MCMC maximum likelihood analysis that takes into account the scatter caused by the uncertainties on the individual measurements, we confirm that the ``ring'' moves faster and is more metal-poor than the control	sample, with the means of the distributions being separated by more than 7$\sigma$ and 3$\sigma$ (see Table~\ref{tab:stat_parameters}). K-S tests between the ``ring'' and the whole control sample yield a negligible (\mbox{p-value $< 10^{-8}$} in velocity) and unlikely (\mbox{p-value $<$ 0.05} in $\Sigma$Mg) probability that they were extracted from the same parent distributions. We note that neither the ``ring'' nor the control sample have a systemic velocity equal to the literature values, while there are not appreciable changes in the velocity dispersion.
	
	\begin{table}	
		\caption{Means and standard deviations of the DRF LOS velocities and $\Sigma$Mg distributions for the ``ring'' and control samples, obtained with sample formulas and the MCMC Hammer.}
		\centering
		\begin{tabular}{ccc}
			\hline
			Parameter & ``Ring'' & Control\\
			& sample & sample\\
			\hline
			$\overline{\textrm{v}}\textsubscript{DRF}$ (km/s), sample formula & 2.79 & -1.77 \\		
			\noalign{\smallskip}
			$\sigma$(v\textsubscript{DRF}) (km/s), sample formula & 8.93 & 8.47 \\
			\noalign{\smallskip}
			$\overline{\textrm{v}}\textsubscript{DRF}$ (km/s), MCMC & 2.93$\substack{+0.65\\-0.65}$ & -1.87$\substack{+0.58\\-0.58}$\\
			\noalign{\smallskip}
			$\sigma$(v\textsubscript{DRF}) (km/s), MCMC & 7.33$\substack{+0.51\\-0.47}$ & 7.27$\substack{+0.46\\-0.43}$\\
			\noalign{\smallskip}
			$\overline{\Sigma\textrm{Mg}}$ (\AA), sample formula & 0.254 & 0.281 \\		
			\noalign{\smallskip}	
			$\sigma$($\Sigma$Mg) (\AA), sample formula & 0.104 & 0.120 \\
			\noalign{\smallskip}
			$\overline{\Sigma\textrm{Mg}}$ (\AA), MCMC & 0.246$\substack{+0.009\\-0.009}$ & 0.273$\substack{+0.009\\-0.009}$\\
			\noalign{\smallskip}
			$\sigma$($\Sigma$Mg) (\AA), MCMC & 0.046$\substack{+0.009\\-0.009}$ & 0.069$\substack{+0.009\\-0.009}$\\
			\hline
		\end{tabular}
		\label{tab:stat_parameters}
	\end{table}
		
	In order to estimate the number of ``ring'' members from its selected candidates, we summed all their membership probabilities according to the Gaussian distributions fitted for the control and ``ring'' samples, obtaining a total of 84 stars following the velocity and $\Sigma$Mg distributions of the ``ring''; that is $\sim20$\% of the total 440 analysed stars. This would set a lower limit in the number of ``ring'' members if considered as an accreted object.
	
	From Fig.~\ref{fig:distributions} it is also apparent that neither the ``ring'' nor the control sample velocity distributions are Gaussians. In order to substantiate this statement, while taking into account the fact that the shape of the distributions might be influenced by the measurement errors, we randomly extracted numerous resamples from both convolved distributions, each with the same number of members as their original samples, and performed an Anderson-Darling test to each of them. As a result, we obtained an unlikely probability for both samples of being extracted from a Gaussian velocity distribution (median \mbox{p-values $<$ 0.007} and 0.05 from the control and ``ring'' resamples, respectively). While this fact in itself does not necessarily imply that the populations are not in dynamical equilibrium, it appears indicative of complex stellar kinematics, probably due to the left-over of the accretion/merging event that produced the ``ring-like'' feature. 
	
\begin{figure}
		\centering
		\includegraphics[trim={2cm 0cm 8.25cm 0.75cm},clip,width=\hsize]{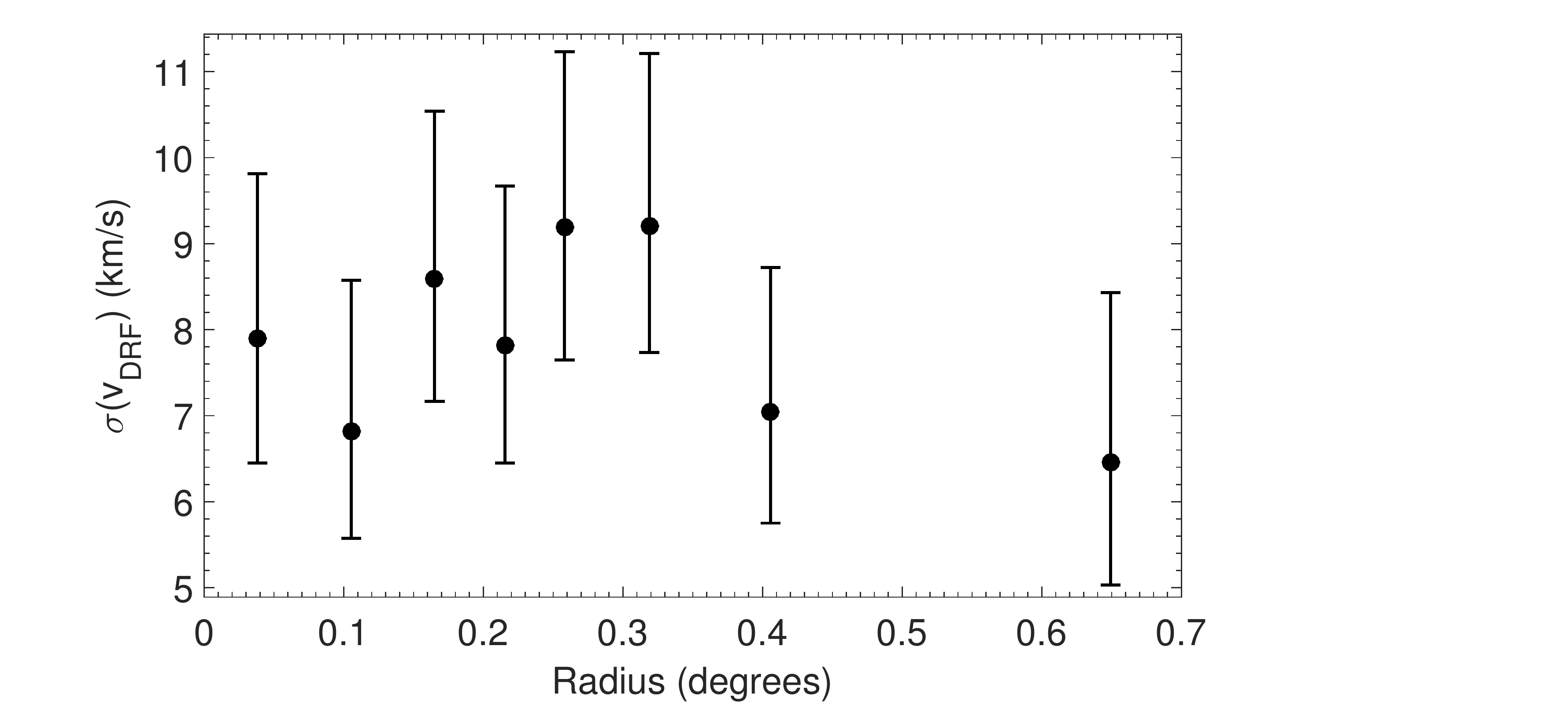}
		\caption{DRF LOS velocity dispersion profile of the blue CMD-selected spectroscopic members as a function of the elliptical radius.}
		\label{fig_disp_vs_radius_blue_sex}
	\end{figure}

	Finally, the middle and bottom panels of Fig.~\ref{fig:kine} show the smoothed DRF velocity and $\Sigma$Mg maps associated to the targets with magnitude and colours that place them in the blue and red CMD selections, respectively, used in Sect.~\ref{sec:photometry}. The round shape observed for the blue population selected in the CTIO/DECam photometric data is reflected in the round shape of the blue spectroscopic targets, as it is the flattened and clumpy distribution of the red population, despite the much brighter magnitude cut used in the spectroscopic sample, just below the HB. In spite of the lower statistics caused by the CMD selection and the colour uncertainties mixing both populations, we can still distinguish the ``ring'' in the blue selection, both in the velocity and $\Sigma$Mg maps, but not in the red selection, in which it coincides with a lack of stars. Therefore, even the population of CMD-blue selected stars, with a seemingly regular spatial distribution, harbors complex kinematic features, whose origin is difficult to understand within a quiet evolutionary history, and would strongly suggest a past accretion or merging event. Being the spatial distribution of the blue stars the most axisymmetric of the CMD-selected components, we show in Fig.~\ref{fig_disp_vs_radius_blue_sex} the LOS velocity dispersion profile of the CMD-selected spectroscopic members in the DRF system as complementary information. It was calculated using the expectation maximization (EM) technique outlined in W09, in the alternative form described in C18. It is possible that the ``ring'' feature is contributing to the slightly larger velocity dispersion values visible between a radius of 0.1 and 0.3 deg.
	
	We note that the cold kinematic substructures previously detected in W06 and K04 fall within the inner region of the control sample (see black crosses in the top-left panel of Fig.~\ref{fig:kine}), while the cold substructure detected in B11 for the metal-poor stars encompasses both parts of the control and ``ring'' samples. As in W06, in this work we do not detect any cold substructure for the metal-poor stars where B11 detected it, that is within a major axis distance of 0.2 deg (Fig.~\ref{fig_disp_vs_radius_blue_sex}). This is probably due to the different number of stars being analyzed in this region (6 in B11 vs. 18 per bin in this work) and to the different selection in metallicity. Regarding the ones detected in W06 and K04, we have not performed an analysis of the local velocity dispersion properties of all the available stars, therefore our findings are not excluding the presence of these features, which, within an accretion/merging scenario, could perhaps be the local clustering of accreted stars close to apocenter.
	
	\section{Discussion and conclusions}\label{sec:con}
	
	By analyzing literature wide-area photometric and spectroscopic datasets for large samples of individual stars in the Sextans dSph, we have uncovered anomalies in both the spatial distribution and kinematic properties of Sextans' stellar component. We have found a clearly different spatial distribution for the blue and red RGB (and MSTO) stars, which is almost round and regular for the former, while being asymmetric and clumpy for the latter. In addition we have found that their different spatial distributions are the responsible of the dependence of the obtained structural parameters on the chosen magnitude cut in C18 and the underdensities there detected. Comparison with isochrones and spectroscopic metallicities showed that the photometrically selected blue stars are on average more metal-poor than the red stars.

    In the analysis of the spectroscopic dataset, we have found that Sextans hosts a ``ring-like'' feature which is detected both in the kinematic and relative metallicity (as given by the $\Sigma$Mg) properties. Despite there being a few wiggles in the LOS velocity dispersion profile of Sextans (W09, B11 and Fig.~\ref{fig_disp_vs_radius_blue_sex}), as the ``ring'' feature can be described as a tilted ellipse, it was not previously detected as a clear change in the mean velocity or increase in the velocity dispersion at a given radius when studying these properties via binning individual velocities in elliptical annuli. The CMD-selected, seemingly regular looking, blue component has actually complex underlying properties, being the composition of at least two velocity and metallicity distributions with different means, one of which forms the aforementioned ``ring''. It has to be noted that both the velocities of the ``ring'' and the ones of the surrounding stars have an unlikely probability of being extracted from a Gaussian distribution, which probably indicates even further kinematic complexities within these populations.	
	
	A likely explanation for the anomalies we detected is a past accretion/merger event; in fact, several of the characteristics are	reminiscent of those seen in on-going accretion events or in systems thought to have experienced an accretion/merger. For example, the clump from the red RGB stars in Sextans resembles a portion of a stellar stream, shell or ripple related to accretion events detected in larger galaxies \citep{martinez-delgado09,martinez-delgado12,rich12}, while ``ring-like'' projected shapes (usually referred as arcing loops or just stellar streams) similar to the one detected in this work have also been found in more massive galaxies \citep{martinez-delgado08,chonis11}.
	
	As in the case of Sextans, the Fornax dSph, which is possibly the result of the late-merger of a bound pair \citep{amorisco12}, shows a shell-like feature within its main-body \citep{coleman04}, a clump off-set from the centre after dividing its RGB into a blue and a red portion \citep[Fig. 8 in][]{battaglia06}, and was found to display hints for a non-equilibrium kinematics in its metal-poor stellar population in the central regions \citep{battaglia06}; later on, an even more complex picture was unveiled, with evidences for counter-rotation between the metal-poor and metal-rich component \citep{amorisco12}. The kinematically detected ``ring'' in Sextans bears a morphological resemblance to the stellar stream in the And~II dSph \citep{amorisco14},	which likely merged with another dwarf of at least one tenth of And~II stellar mass. Differently from the case of And~II though, where a clear velocity gradient is observed across the stream, the ``ring'' in Sextans appears to have overall a higher systemic LOS velocity than the rest	of Sextans population, perhaps resulting from an helical orbit. However the ``ring'' could be telling just part of the story and being formed by the dispersed stars of the accreted object which happen to have a more distinct kinematic pattern, while the rest is {\it hiding} under a more normal kinematics, and showing up in the non-Gaussianity of the LOS velocity distribution of stars at the location of the ``ring'' and outside.
	
	Since the kinematically detected ``ring'' displays on average a lower metallicity than the rest of the population, it is possible that it is, at least partly, formed by the remnant of an object more metal-poor than Sextans; this appears reasonable at the light of the luminosity-metallicity relation of LG dwarfs \citep{kirby13}. But how did the asymmetric, clumpy, redder/more metal-rich component originate? Did it originate {\it in-situ} during the accretion or merger event? Simulations such as those by \cite{benitez-llambay16} point to the possibility that the old stellar component of dwarf galaxies might be assembled in mergers, while the younger one would mostly form {\it in situ} from the accreted gas a few Gyr after. However, the star formation history and general chemical properties of Sextans are compatible with the galaxy being a fossil from the pre-reionization era \citep{bettinelli18,revaz18}, which imposes tight constraints for the sequence of events that shaped the object and makes it hard to understand how the clumps and asymmetries could have survived for such a long time. Clearly, understanding the details of the possible accretion/merger event experienced by Sextans will require dedicated simulations, which should take into account the wealth of observational information available for this galaxy.
		
	Finally, it is worth mentioning that while the smallest galaxy with univocal signs of accretion up to date was the And~II dSph \citep{amorisco14}, which has a stellar mass of $\sim10^{7} M_{\odot}$ \citep{mcconnachie07}, Sextans has a stellar mass of just $\sim5\times10^{5} M_{\odot}$ \citep{mcconnachie12}, that is to say $\sim5$\% of the luminous mass of And~II. This places Sextans as the smallest galaxy presenting clear observational signs of accretion to date.
	
	\section*{Acknowledgements}

    The authors would like to thank M. Irwin, R. Ibata and C. Gallart for useful discussions in the early stages of this work, and the anonymous referee for helpful comments. L. Cicuéndez Salazar acknowledges Fundación la Caixa for the financial support received in the form of a PhD contract. G. Battaglia acknowledges financial support by the Spanish Ministry of Economy and Competitiveness (MINECO) under the Ramón y Cajal Programme (RYC-2012-11537) and the grant AYA2014-56795-P.
	
	Based on observations at Cerro Tololo Inter-American Observatory, National Optical Astronomy Observatory (NOAO Prop. ID: 2015A/1013; PI: B. McMonigal), which is operated by the Association of Universities for Research in Astronomy (AURA) under a coope rative agreement with the National Science Foundation.	
	\begin{quotation}
		This project used data obtained with the Dark Energy Camera (DECam), which was constructed by the Dark Energy Survey (DES) collaboration.

		Funding for the DES Projects has been provided by the DOE and NSF (USA), MISE (Spain), STFC (UK), HEFCE (UK), NCSA (UIUC), KICP (U. Chicago), CCAPP (Ohio State), MIFPA (Texas A\&M), CNPQ, FAPERJ, FINEP (Brazil), MINECO (Spain), DFG (Germany) and the collaborating institutions in the Dark Energy Survey, which are Argonne Lab, UC Santa Cruz, University of Cambridge, CIEMAT-Madrid, University of Chicago, University College London, DES-Brazil Consortium, University of Edinburgh, ETH Z{\"u}rich, Fermilab, University of Illinois, ICE (IEEC-CSIC), IFAE Barcelona, Lawrence Berkeley Lab, LMU M{\"u}nchen and the associated Excellence Cluster Universe, University of Michigan, NOAO, University of Nottingham, Ohio State University, University of Pennsylvania, University of Portsmouth, SLAC National Lab, Stanford University, University of Sussex, and Texas A\&M University.
	\end{quotation}
	
	
	\bibliographystyle{mnras}
	\bibliography{biblio} 

\begin{thebibliography}{}
\makeatletter
\relax
\def\mn@urlcharsother{\let\do\@makeother \do\$\do\&\do\#\do\^\do\_\do\%\do\~}
\def\mn@doi{\begingroup\mn@urlcharsother \@ifnextchar [ {\mn@doi@}
  {\mn@doi@[]}}
\def\mn@doi@[#1]#2{\def\@tempa{#1}\ifx\@tempa\@empty \href
  {http://dx.doi.org/#2} {doi:#2}\else \href {http://dx.doi.org/#2} {#1}\fi
  \endgroup}
\def\mn@eprint#1#2{\mn@eprint@#1:#2::\@nil}
\def\mn@eprint@arXiv#1{\href {http://arxiv.org/abs/#1} {{\tt arXiv:#1}}}
\def\mn@eprint@dblp#1{\href {http://dblp.uni-trier.de/rec/bibtex/#1.xml}
  {dblp:#1}}
\def\mn@eprint@#1:#2:#3:#4\@nil{\def\@tempa {#1}\def\@tempb {#2}\def\@tempc
  {#3}\ifx \@tempc \@empty \let \@tempc \@tempb \let \@tempb \@tempa \fi \ifx
  \@tempb \@empty \def\@tempb {arXiv}\fi \@ifundefined
  {mn@eprint@\@tempb}{\@tempb:\@tempc}{\expandafter \expandafter \csname
  mn@eprint@\@tempb\endcsname \expandafter{\@tempc}}}

\bibitem[\protect\citeauthoryear{{Amorisco} \& {Evans}}{{Amorisco} \&
  {Evans}}{2012}]{amorisco12}
{Amorisco} N.~C.,  {Evans} N.~W.,  2012, \mn@doi [\apjl]
  {10.1088/2041-8205/756/1/L2}, \href
  {http://cdsads.u-strasbg.fr/abs/2012ApJ...756L...2A} {756, L2}

\bibitem[\protect\citeauthoryear{{Amorisco}, {Evans}  \& {van de
  Ven}}{{Amorisco} et~al.}{2014}]{amorisco14}
{Amorisco} N.~C.,  {Evans} N.~W.,   {van de Ven} G.,  2014, \mn@doi [\nat]
  {10.1038/nature12995}, \href
  {http://adsabs.harvard.edu/abs/2014Natur.507..335A} {507, 335}

\bibitem[\protect\citeauthoryear{{Battaglia}}{{Battaglia}}{2007}]{battaglia07}
{Battaglia} G.,  2007, PhD thesis, Kapteyn Astronomical Institute, University
  of Groningen

\bibitem[\protect\citeauthoryear{{Battaglia} et~al.,}{{Battaglia}
  et~al.}{2006}]{battaglia06}
{Battaglia} G.,  et~al., 2006, \mn@doi [\aap] {10.1051/0004-6361:20065720},
  \href {http://adsabs.harvard.edu/abs/2006A%26A...459..423B} {459, 423}

\bibitem[\protect\citeauthoryear{{Battaglia}, {Tolstoy}, {Helmi}, {Irwin},
  {Parisi}, {Hill}  \& {Jablonka}}{{Battaglia} et~al.}{2011}]{battaglia11}
{Battaglia} G.,  {Tolstoy} E.,  {Helmi} A.,  {Irwin} M.,  {Parisi} P.,  {Hill}
  V.,   {Jablonka} P.,  2011, \mn@doi [\mnras]
  {10.1111/j.1365-2966.2010.17745.x}, \href
  {http://adsabs.harvard.edu/abs/2011MNRAS.411.1013B} {411, 1013}

\bibitem[\protect\citeauthoryear{{Bellazzini}, {Ferraro}, {Origlia}, {Pancino},
  {Monaco}  \& {Oliva}}{{Bellazzini} et~al.}{2002}]{bellazzini02}
{Bellazzini} M.,  {Ferraro} F.~R.,  {Origlia} L.,  {Pancino} E.,  {Monaco} L.,
   {Oliva} E.,  2002, \mn@doi [\aj] {10.1086/344794}, \href
  {http://adsabs.harvard.edu/abs/2002AJ....124.3222B} {124, 3222}

\bibitem[\protect\citeauthoryear{{Ben{\'{\i}}tez-Llambay}, {Navarro}, {Abadi},
  {Gottl{\"o}ber}, {Yepes}, {Hoffman}  \& {Steinmetz}}{{Ben{\'{\i}}tez-Llambay}
  et~al.}{2016}]{benitez-llambay16}
{Ben{\'{\i}}tez-Llambay} A.,  {Navarro} J.~F.,  {Abadi} M.~G.,  {Gottl{\"o}ber}
  S.,  {Yepes} G.,  {Hoffman} Y.,   {Steinmetz} M.,  2016, \mn@doi [\mnras]
  {10.1093/mnras/stv2722}, \href
  {http://adsabs.harvard.edu/abs/2016MNRAS.456.1185B} {456, 1185}

\bibitem[\protect\citeauthoryear{{Bettinelli}, {Hidalgo}, {Cassisi}, {Aparicio}
   \& {Piotto}}{{Bettinelli} et~al.}{2018}]{bettinelli18}
{Bettinelli} M.,  {Hidalgo} S.~L.,  {Cassisi} S.,  {Aparicio} A.,   {Piotto}
  G.,  2018, \mn@doi [\mnras] {10.1093/mnras/sty226}, \href
  {http://adsabs.harvard.edu/abs/2018MNRAS.tmp..226B} {}

\bibitem[\protect\citeauthoryear{{Casetti-Dinescu}, {Girard}  \&
  {Schriefer}}{{Casetti-Dinescu} et~al.}{2017}]{casetti17}
{Casetti-Dinescu} D.~I.,  {Girard} T.~M.,   {Schriefer} M.,  2017, preprint,
  \href {http://adsabs.harvard.edu/abs/2017arXiv171002462C} {} (\mn@eprint
  {arXiv} {1710.02462})

\bibitem[\protect\citeauthoryear{{Chonis}, {Mart{\'{\i}}nez-Delgado}, {Gabany},
  {Majewski}, {Hill}, {Gralak}  \& {Trujillo}}{{Chonis}
  et~al.}{2011}]{chonis11}
{Chonis} T.~S.,  {Mart{\'{\i}}nez-Delgado} D.,  {Gabany} R.~J.,  {Majewski}
  S.~R.,  {Hill} G.~J.,  {Gralak} R.,   {Trujillo} I.,  2011, \mn@doi [\aj]
  {10.1088/0004-6256/142/5/166}, \href
  {http://adsabs.harvard.edu/abs/2011AJ....142..166C} {142, 166}

\bibitem[\protect\citeauthoryear{{Cicu{\'e}ndez} et~al.,}{{Cicu{\'e}ndez}
  et~al.}{2018}]{yo18}
{Cicu{\'e}ndez} L.,  et~al., 2018, \mn@doi [\aap]
  {10.1051/0004-6361/201731450}, \href
  {http://adsabs.harvard.edu/abs/2018A%26A...609A..53C} {609, A53} (C18)

\bibitem[\protect\citeauthoryear{{Coleman}, {Da Costa}, {Bland-Hawthorn},
  {Mart{\'{\i}}nez-Delgado}, {Freeman}  \& {Malin}}{{Coleman}
  et~al.}{2004}]{coleman04}
{Coleman} M.,  {Da Costa} G.~S.,  {Bland-Hawthorn} J.,
  {Mart{\'{\i}}nez-Delgado} D.,  {Freeman} K.~C.,   {Malin} D.,  2004, \mn@doi
  [\aj] {10.1086/381298}, \href
  {http://adsabs.harvard.edu/abs/2004AJ....127..832C} {127, 832}

\bibitem[\protect\citeauthoryear{{Deason}, {Wetzel}  \&
  {Garrison-Kimmel}}{{Deason} et~al.}{2014}]{deason14}
{Deason} A.,  {Wetzel} A.,   {Garrison-Kimmel} S.,  2014, \mn@doi [\apj]
  {10.1088/0004-637X/794/2/115}, \href
  {http://adsabs.harvard.edu/abs/2014ApJ...794..115D} {794, 115}

\bibitem[\protect\citeauthoryear{{Ebrov{\'a}} \& {{\L}okas}}{{Ebrov{\'a}} \&
  {{\L}okas}}{2015}]{ebrova15}
{Ebrov{\'a}} I.,  {{\L}okas} E.~L.,  2015, \mn@doi [\apj]
  {10.1088/0004-637X/813/1/10}, \href
  {http://adsabs.harvard.edu/abs/2015ApJ...813...10E} {813, 10}

\bibitem[\protect\citeauthoryear{{Fakhouri}, {Ma}  \&
  {Boylan-Kolchin}}{{Fakhouri} et~al.}{2010}]{fakhouri10}
{Fakhouri} O.,  {Ma} C.-P.,   {Boylan-Kolchin} M.,  2010, \mn@doi [\mnras]
  {10.1111/j.1365-2966.2010.16859.x}, \href
  {http://adsabs.harvard.edu/abs/2010MNRAS.406.2267F} {406, 2267}

\bibitem[\protect\citeauthoryear{{Foreman-Mackey}, {Hogg}, {Lang}  \&
  {Goodman}}{{Foreman-Mackey} et~al.}{2013}]{foreman-mackey13}
{Foreman-Mackey} D.,  {Hogg} D.~W.,  {Lang} D.,   {Goodman} J.,  2013, \mn@doi
  [\pasp] {10.1086/670067}, \href
  {http://adsabs.harvard.edu/abs/2013PASP..125..306F} {125, 306}

\bibitem[\protect\citeauthoryear{{Fouquet}, {{\L}okas}, {del Pino}  \&
  {Ebrov{\'a}}}{{Fouquet} et~al.}{2017}]{fouquet17}
{Fouquet} S.,  {{\L}okas} E.~L.,  {del Pino} A.,   {Ebrov{\'a}} I.,  2017,
  \mn@doi [\mnras] {10.1093/mnras/stw2510}, \href
  {http://adsabs.harvard.edu/abs/2017MNRAS.464.2717F} {464, 2717}

\bibitem[\protect\citeauthoryear{{Gaia Collaboration} et~al.,}{{Gaia
  Collaboration} et~al.}{2018}]{helmi18}
{Gaia Collaboration} et~al., 2018, preprint, \href
  {http://adsabs.harvard.edu/abs/2018arXiv180409381G} {} (\mn@eprint {arXiv}
  {1804.09381})

\bibitem[\protect\citeauthoryear{{Ho} et~al.,}{{Ho} et~al.}{2012}]{ho12}
{Ho} N.,  et~al., 2012, \mn@doi [\apj] {10.1088/0004-637X/758/2/124}, \href
  {http://adsabs.harvard.edu/abs/2012ApJ...758..124H} {758, 124}

\bibitem[\protect\citeauthoryear{{Kacharov} et~al.,}{{Kacharov}
  et~al.}{2017}]{kacharov17}
{Kacharov} N.,  et~al., 2017, \mn@doi [\mnras] {10.1093/mnras/stw3188}, \href
  {http://adsabs.harvard.edu/abs/2017MNRAS.466.2006K} {466, 2006}

\bibitem[\protect\citeauthoryear{{Kirby} et~al.,}{{Kirby}
  et~al.}{2010}]{kirby10}
{Kirby} E.~N.,  et~al., 2010, \mn@doi [\apjs] {10.1088/0067-0049/191/2/352},
  \href {http://adsabs.harvard.edu/abs/2010ApJS..191..352K} {191, 352}

\bibitem[\protect\citeauthoryear{{Kirby}, {Cohen}, {Guhathakurta}, {Cheng},
  {Bullock}  \& {Gallazzi}}{{Kirby} et~al.}{2013}]{kirby13}
{Kirby} E.~N.,  {Cohen} J.~G.,  {Guhathakurta} P.,  {Cheng} L.,  {Bullock}
  J.~S.,   {Gallazzi} A.,  2013, \mn@doi [\apj] {10.1088/0004-637X/779/2/102},
  \href {http://adsabs.harvard.edu/abs/2013ApJ...779..102K} {779, 102}

\bibitem[\protect\citeauthoryear{{Kleyna}, {Wilkinson}, {Gilmore}  \&
  {Evans}}{{Kleyna} et~al.}{2003}]{kleyna03}
{Kleyna} J.~T.,  {Wilkinson} M.~I.,  {Gilmore} G.,   {Evans} N.~W.,  2003,
  \mn@doi [\apjl] {10.1086/375522}, \href
  {http://adsabs.harvard.edu/abs/2003ApJ...588L..21K} {588, L21}

\bibitem[\protect\citeauthoryear{{Kleyna}, {Wilkinson}, {Evans}  \&
  {Gilmore}}{{Kleyna} et~al.}{2004}]{kleyna04}
{Kleyna} J.~T.,  {Wilkinson} M.~I.,  {Evans} N.~W.,   {Gilmore} G.,  2004,
  \mn@doi [\mnras] {10.1111/j.1365-2966.2004.08434.x}, \href
  {http://adsabs.harvard.edu/abs/2004MNRAS.354L..66K} {354, L66}

\bibitem[\protect\citeauthoryear{{Mart{\'{\i}}nez-Delgado}, {Pe{\~n}arrubia},
  {Gabany}, {Trujillo}, {Majewski}  \& {Pohlen}}{{Mart{\'{\i}}nez-Delgado}
  et~al.}{2008}]{martinez-delgado08}
{Mart{\'{\i}}nez-Delgado} D.,  {Pe{\~n}arrubia} J.,  {Gabany} R.~J.,
  {Trujillo} I.,  {Majewski} S.~R.,   {Pohlen} M.,  2008, \mn@doi [\apj]
  {10.1086/592555}, \href {http://adsabs.harvard.edu/abs/2008ApJ...689..184M}
  {689, 184}

\bibitem[\protect\citeauthoryear{{Mart{\'{\i}}nez-Delgado}, {Pohlen}, {Gabany},
  {Majewski}, {Pe{\~n}arrubia}  \& {Palma}}{{Mart{\'{\i}}nez-Delgado}
  et~al.}{2009}]{martinez-delgado09}
{Mart{\'{\i}}nez-Delgado} D.,  {Pohlen} M.,  {Gabany} R.~J.,  {Majewski} S.~R.,
   {Pe{\~n}arrubia} J.,   {Palma} C.,  2009, \mn@doi [\apj]
  {10.1088/0004-637X/692/2/955}, \href
  {http://adsabs.harvard.edu/abs/2009ApJ...692..955M} {692, 955}

\bibitem[\protect\citeauthoryear{{Mart{\'{\i}}nez-Delgado}
  et~al.,}{{Mart{\'{\i}}nez-Delgado} et~al.}{2012}]{martinez-delgado12}
{Mart{\'{\i}}nez-Delgado} D.,  et~al., 2012, \mn@doi [\apjl]
  {10.1088/2041-8205/748/2/L24}, \href
  {http://adsabs.harvard.edu/abs/2012ApJ...748L..24M} {748, L24}

\bibitem[\protect\citeauthoryear{{McConnachie}}{{McConnachie}}{2012}]{mcconnachie12}
{McConnachie} A.~W.,  2012, \mn@doi [\aj] {10.1088/0004-6256/144/1/4}, \href
  {http://adsabs.harvard.edu/abs/2012AJ....144....4M} {144, 4}

\bibitem[\protect\citeauthoryear{{McConnachie}, {Arimoto}  \&
  {Irwin}}{{McConnachie} et~al.}{2007}]{mcconnachie07}
{McConnachie} A.~W.,  {Arimoto} N.,   {Irwin} M.,  2007, \mn@doi [\mnras]
  {10.1111/j.1365-2966.2007.11969.x}, \href
  {http://adsabs.harvard.edu/abs/2007MNRAS.379..379M} {379, 379}

\bibitem[\protect\citeauthoryear{{Pace}, {Martinez}, {Kaplinghat}  \&
  {Mu{\~n}oz}}{{Pace} et~al.}{2014}]{pace14}
{Pace} A.~B.,  {Martinez} G.~D.,  {Kaplinghat} M.,   {Mu{\~n}oz} R.~R.,  2014,
  \mn@doi [\mnras] {10.1093/mnras/stu938}, \href
  {http://adsabs.harvard.edu/abs/2014MNRAS.442.1718P} {442, 1718}

\bibitem[\protect\citeauthoryear{{Piatti}}{{Piatti}}{2018}]{piatti18}
{Piatti} A.~E.,  2018, \mn@doi [\mnras] {10.1093/mnras/stx2471}, \href
  {http://adsabs.harvard.edu/abs/2018MNRAS.473..492P} {473, 492}

\bibitem[\protect\citeauthoryear{{Revaz} \& {Jablonka}}{{Revaz} \&
  {Jablonka}}{2018}]{revaz18}
{Revaz} Y.,  {Jablonka} P.,  2018, preprint, \href
  {http://adsabs.harvard.edu/abs/2018arXiv180106222R} {} (\mn@eprint {arXiv}
  {1801.06222})

\bibitem[\protect\citeauthoryear{{Rich}, {Collins}, {Black}, {Longstaff},
  {Koch}, {Benson}  \& {Reitzel}}{{Rich} et~al.}{2012}]{rich12}
{Rich} R.~M.,  {Collins} M.~L.~M.,  {Black} C.~M.,  {Longstaff} F.~A.,  {Koch}
  A.,  {Benson} A.,   {Reitzel} D.~B.,  2012, \mn@doi [\nat]
  {10.1038/nature10837}, \href
  {http://adsabs.harvard.edu/abs/2012Natur.482..192R} {482, 192}

\bibitem[\protect\citeauthoryear{{Richardson} et~al.,}{{Richardson}
  et~al.}{2011}]{richardson11}
{Richardson} J.~C.,  et~al., 2011, \mn@doi [\apj] {10.1088/0004-637X/732/2/76},
  \href {http://adsabs.harvard.edu/abs/2011ApJ...732...76R} {732, 76}

\bibitem[\protect\citeauthoryear{{Walker}, {Mateo}, {Olszewski}, {Pal}, {Sen}
  \& {Woodroofe}}{{Walker} et~al.}{2006}]{walker06}
{Walker} M.~G.,  {Mateo} M.,  {Olszewski} E.~W.,  {Pal} J.~K.,  {Sen} B.,
  {Woodroofe} M.,  2006, \mn@doi [\apjl] {10.1086/504522}, \href
  {http://adsabs.harvard.edu/abs/2006ApJ...642L..41W} {642, L41}

\bibitem[\protect\citeauthoryear{{Walker}, {Mateo}  \& {Olszewski}}{{Walker}
  et~al.}{2008}]{walker08}
{Walker} M.~G.,  {Mateo} M.,   {Olszewski} E.~W.,  2008, \mn@doi [\apjl]
  {10.1086/595586}, \href {http://adsabs.harvard.edu/abs/2008ApJ...688L..75W}
  {688, L75}

\bibitem[\protect\citeauthoryear{{Walker}, {Mateo}  \& {Olszewski}}{{Walker}
  et~al.}{2009a}]{walker09c}
{Walker} M.~G.,  {Mateo} M.,   {Olszewski} E.~W.,  2009a, \mn@doi [\aj]
  {10.1088/0004-6256/137/2/3100}, \href
  {http://adsabs.harvard.edu/abs/2009AJ....137.3100W} {137, 3100}

\bibitem[\protect\citeauthoryear{{Walker}, {Mateo}, {Olszewski}, {Sen}  \&
  {Woodroofe}}{{Walker} et~al.}{2009b}]{walker09}
{Walker} M.~G.,  {Mateo} M.,  {Olszewski} E.~W.,  {Sen} B.,   {Woodroofe} M.,
  2009b, \mn@doi [\aj] {10.1088/0004-6256/137/2/3109}, \href
  {http://adsabs.harvard.edu/abs/2009AJ....137.3109W} {137, 3109}

\bibitem[\protect\citeauthoryear{{White} \& {Rees}}{{White} \&
  {Rees}}{1978}]{white78}
{White} S.~D.~M.,  {Rees} M.~J.,  1978, \mn@doi [\mnras]
  {10.1093/mnras/183.3.341}, \href
  {http://adsabs.harvard.edu/abs/1978MNRAS.183..341W} {183, 341}

\bibitem[\protect\citeauthoryear{{Wolf}, {Martinez}, {Bullock}, {Kaplinghat},
  {Geha}, {Mu{\~n}oz}, {Simon}  \& {Avedo}}{{Wolf} et~al.}{2010}]{wolf10}
{Wolf} J.,  {Martinez} G.~D.,  {Bullock} J.~S.,  {Kaplinghat} M.,  {Geha} M.,
  {Mu{\~n}oz} R.~R.,  {Simon} J.~D.,   {Avedo} F.~F.,  2010, \mn@doi [\mnras]
  {10.1111/j.1365-2966.2010.16753.x}, \href
  {http://adsabs.harvard.edu/abs/2010MNRAS.406.1220W} {406, 1220}

\makeatother
\end{thebibliography}
	
	
	\appendix
	
	\section{Decontamination of 2D maps} \label{sec:app}
	
	The method we developed to decontaminate the 2D spatial distribution of individual sources in the bRGB and rRGB selection boxes without spatially binning the data was inspired by the one developed in C18 to decontaminate the radial cumulative distribution functions (CDFs) of the different Sextans' evolutionary phases.
	
	We first calculate the ratio between the number of stars falling in the CMD selection box of interest but located well away from the galaxy centre, for example outside the King tidal radius, therefore at a spatial location expected to be occupied only by contaminants, and the number of objects in this same spatial region but in a part of CMD known not to contain stars from the target galaxy; in this case we used the colour cut $g-r>1$, as in C18. This ratio reflects the expected proportion of contaminants in the CMD selection box of interest with respect to the red ones ($g-r>1$).
	
	The algorithm next randomly resamples all the red contaminants according to this ratio and decontaminates the two-dimensional spatial distribution as we did with the one-dimensional radial CDFs in C18, that is by removing those stars spatially closest to the resampled contaminants. We implemented two other features in this decontamination method. On one hand, in order to prevent removing some stars with too large separation from their closest red contaminant, we compute the probability distribution function of the separations between the stars from the sample and their closest resampled red contaminants and limit the maximum allowed separation to be the percentile 99\% of this probability distribution. On the other hand, inspired by the procedure developed in \cite{piatti18} to calculate stellar membership probabilities in a globular cluster, instead of making an unique decontamination from just one resample of the red contaminants, our algorithm generates numerous decontaminated maps (i.e. 1000) from many different random resamplings of the red contaminants. Then, the probability that a star belongs to the galaxy is the ratio between the number of times that it is not removed and the number of decontaminations performed. We found that the contaminants start being negligible when plotting the spatial locations of those stars with a membership probability larger than $\sim 60$-70\%.
	
	
	\bsp	
	\label{lastpage}
\end{document}